\PassOptionsToPackage{dvipsnames,table}{xcolor}
\documentclass[sigconf]{acmart}

\usepackage{amsmath}
\usepackage{bm}
\usepackage{amsfonts}
\usepackage{mathtools}
\usepackage{tikz}
\usetikzlibrary{arrows.meta}
\usepackage{xspace}
\usepackage{subfig}
\usepackage{pgfplots}
\pgfplotsset{compat=1.12}
\usepgfplotslibrary{groupplots}
\usepackage{boldline}
\usepackage[subtle]{savetrees}
\usepackage{autobreak}

\pgfplotsset{select coords between index/.style 2 args={
    x filter/.code={
        \ifnum\coordindex<#1\fi
        \ifnum\coordindex>#2\fi
    }
}}

\def\kg{$\mathcal{KG}$\xspace}
\def\kgs{$\mathcal{KG}s$\xspace}

\newcommand{\myrowcolour}{\rowcolor[gray]{0.925}}

\newenvironment{customlegend}[1][]{%
    \begingroup
    \csname pgfplots@init@cleared@structures\endcsname
    \pgfplotsset{#1}%
}{%
    \csname pgfplots@createlegend\endcsname
    \endgroup
}%
\def\addlegendimage{\csname pgfplots@addlegendimage\endcsname}

\def \framework{KGFlex\xspace}
\def \frameworkwithoutspace{KGFlex}

\theoremstyle{definition}
\newtheorem*{myexp*}{Example}

\newenvironment{rexp}
{\begin{myexp*}{\textit{\textbf{{(continued)}}}}}
  {\end{myexp*}}

\makeatletter
\newcommand\feature[1]{%
  \@tempcnta=0
  \langle
  \@for\@ii:=#1\do{%
    \@insertbreakingcomma
    \textit{\@ii}
  }%
  \rangle
}
\def\@insertbreakingcomma{%
  \ifnum \@tempcnta = 0 \else, \linebreak[1] \fi
  \advance\@tempcnta\@ne
}
\makeatother

\AtBeginDocument{%
  \providecommand\BibTeX{{%
    \normalfont B\kern-0.5em{\scshape i\kern-0.25em b}\kern-0.8em\TeX}}}

\copyrightyear{2021} 
\acmYear{2021} 
\setcopyright{acmcopyright}\acmConference[RecSys '21]{Fifteenth ACM Conference on Recommender Systems}{September 27-October 1, 2021}{Amsterdam, Netherlands}
\acmBooktitle{Fifteenth ACM Conference on Recommender Systems (RecSys '21), September 27-October 1, 2021, Amsterdam, Netherlands}
\acmPrice{15.00}
\acmDOI{10.1145/3460231.3474243}
\acmISBN{978-1-4503-8458-2/21/09}

\begin{document}

\title[Sparse Feature Factorization for Recommender Systems with KGs]{Sparse Feature Factorization for Recommender Systems with Knowledge Graphs}

\author{Vito Walter Anelli}
\authornote{Authors are listed in alphabetical order. Corresponding authors: A. Ferrara (\url{antonio.ferrara@poliba.it}) and V. W. Anelli (\url{vitowalter.anelli@poliba.it}).}
\email{vitowalter.anelli@poliba.it}
\affiliation{%
  \institution{Polytechnic University of Bari}
  \city{Bari}
  \country{Italy}
}

\author{Tommaso Di Noia}
\email{tommaso.dinoia@poliba.it}
\affiliation{%
  \institution{Polytechnic University of Bari}
  \city{Bari}
  \country{Italy}
}

\author{Eugenio Di Sciascio}
\email{eugenio.disciascio@poliba.it}
\affiliation{%
  \institution{Polytechnic University of Bari}
  \city{Bari}
  \country{Italy}
}

\author{Antonio Ferrara}
\authornotemark[1]
\email{antonio.ferrara@poliba.it}
\affiliation{%
  \institution{Polytechnic University of Bari}
  \city{Bari}
  \country{Italy}
}

\author{Alberto Carlo Maria Mancino}
\email{alberto.mancino@poliba.it}
\affiliation{%
  \institution{Polytechnic University of Bari}
  \city{Bari}
  \country{Italy}
}

\renewcommand{\shortauthors}{Anelli et al.}

\begin{abstract}
Deep Learning and factorization-based collaborative filtering recommendation models have undoubtedly dominated the scene of recommender systems in recent years. 
However, despite their outstanding performance, these methods require a training time proportional to the size of the embeddings and it further increases when also side information is considered for the computation of the recommendation list.
In fact, in these cases we have that with a large number of high-quality features, the resulting models are more complex and difficult to train.
This paper addresses this problem by presenting \framework: a sparse factorization approach that grants an even greater degree of expressiveness.
To achieve this result, \framework analyzes the historical data to understand the dimensions the user decisions depend on (e.g., movie direction, musical genre, nationality of book writer).
\framework represents each item feature as an embedding and it models user-item interactions as a factorized entropy-driven combination of the item attributes relevant to the user. \framework facilitates the training process by letting users update only those relevant features on which they base their decisions.
In other words, the user-item prediction is mediated by the user's personal view that considers only relevant features.
An extensive experimental evaluation shows the approach's effectiveness, considering the recommendation results' accuracy, diversity, and induced bias.
The public implementation of \framework is available at \url{https://split.to/kgflex}.
\end{abstract}

\ccsdesc[500]{Information systems~Recommender systems}
\ccsdesc[500]{Information systems~Personalization}

\keywords{feature factorization, entropy, knowledge graphs}

\maketitle

\section{Introduction}
The history of automated recommendation is closely linked to the evolution of collaborative filtering techniques.
Their notable accuracy has unquestionably helped Recommender Systems getting famous.
Despite their leading performance, these methods are based on the simple idea to recommend certain items since "similar users have experienced those items", or "other users, who have experienced the same items, have also experienced those items."
In the past, Matrix Factorization~\cite{DBLP:journals/computer/KorenBV09} and Nearest Neighbors were the main algorithms to implement Collaborative Filtering and, over the last years, Deep Learning models~\cite{DBLP:conf/uic/ChakrabortyTRHA17} have joined this shortlist.
The main limitation of these approaches is the requirement of many parameters that further increase at least proportionally according to the dataset size.

Differently from collaborative approaches, content-based recommendation techniques aim to identify the common characteristics of items that a user liked in the past~\cite{DBLP:conf/adaptive/PazzaniB07}. 
They match the user profile against the attributes of the items and recommend new items that share the same features. 
On the one hand, the use of content features can make the model interpretable~\cite{DBLP:journals/corr/abs-1804-11192} while, on the other hand, these techniques suffer from overspecialization since they fail to recommend items that are different from the items enjoyed in the past.
In order to get the benefits of the two approaches and mitigate their drawbacks, researchers worked to integrate into Collaborative Filtering the side information used in content-based approaches such as tags~\cite{DBLP:journals/ijon/ZhuGTLCH16}, images~\cite{DBLP:conf/sigir/AnelliDNMM21}, demographic data~\cite{DBLP:journals/kais/ZhaoLHWWL16}, structured knowledge~\cite{DBLP:conf/semweb/AnelliNSRT19}. 
However, even there, the predominant adoption of large and dense models implies that user-item interactions are predicted by taking into account hundreds or thousands of features.

In this work, we introduce \framework, a knowledge-aware recommendation system, that tackles this issue by exploiting a sparse embedding model with an even greater degree of expressiveness. 
\framework extracts facts and knowledge from publicly available knowledge graphs to describe the catalog items. 
Then, low-dimensionality embeddings are adopted to represent the semantic item features.
\framework models the user-item interaction by combining the subset of item features relevant to the user. 
Moreover, it analyses the user-specific decision-making process of consuming or not consuming an item.
According to that process, the system weights feature embeddings using an entropy-based strategy. 
Therefore, \framework computes, for each user, a set of features their decisions are based on.
According to a principle of expertise, during training, only the features the specific user is expert about are updated for a given user-item pair.
Hence, the user profile itself only contains a personal representation of each relevant feature.

To evaluate the performance of \framework, we conduct extensive experiments on three different publicly available datasets. The content-based features have been extracted from data encoded in the DBpedia\footnote{\url{http://dbpedia.org}} knowledge graph, thanks to public mappings from the dataset items to DBpedia URIs\footnote{\url{https://github.com/sisinflab/LinkedDatasets}}. We evaluate the accuracy and diversity of recommendation results and analyze whether the algorithm produces biased recommendations.
Finally, we study how users' decision-making process differs from \framework's one by graphically showing the semantic shift produced in the recommendation. 
The results show that \framework has competitive accuracy performance, and at the same time, generates highly diversified recommendations with a low induced bias.

\section{Background}
\subsection{Knowledge-aware Recommender Systems (KaRSs)}
Nowadays, modern RSs exploit various side information such as metadata (e.g., tags, reviews)~\cite{DBLP:conf/recsys/NingK12}, social connections~\cite{DBLP:conf/wsdm/BackstromL11}, images~\cite{DBLP:conf/sigir/AnelliDNMM21}, and users-items contextual data~\cite{DBLP:conf/um/AnelliBNBTS17} to build more in-domain~\cite{9216015} (i.e., domain-dependent), cross-domain~\cite{DBLP:journals/umuai/Fernandez-Tobias19}, or context-aware~\cite{DBLP:journals/isci/HuoWNCZ20,DBLP:conf/esws/HildebrandtSMJM19} recommendation models.
Among the diverse information sources, what is, likely, the most relevant source is Knowledge Graphs (\kgs).
Thanks to the heterogeneous domains that \kgs cover, the design of knowledge-based recommendation systems has arisen as a specific research field of its own in the community of RSs, usually referred to by Knowledge-aware Recommender Systems (KaRS~\cite{DBLP:conf/cikm/AnelliN19,DBLP:conf/recsys/AnelliBBNLMNZ18}).
The adoption of \kgs as a source of side-information has generated several advancements in the tasks of recommendation~\cite{DBLP:conf/semweb/AnelliNSRT19}, knowledge completion~\cite{DBLP:conf/www/HeLZLW20}, preference elicitation~\cite{DBLP:journals/semweb/AnelliLNLR20}, user modeling~\cite{DBLP:journals/tois/WangZWZLXG19}, and thus produced a vast literature.
In recent years, the Knowledge-aware Recommender Systems have been particularly impactful for several recommendation tasks: 
\textbf{hybrid collaborative/content-based recommendation}~\cite{DBLP:conf/IEEEwisa/LiXTZT20,DBLP:conf/semweb/AnelliNSRT19}, exploiting the \kg information to suffice the lack of collaborative information and to improve the performance;
\textbf{knowledge-transfer, cross-domain recommendation}~\cite{DBLP:conf/ijcnn/ZhangH0019,DBLP:journals/umuai/Fernandez-Tobias19,DBLP:conf/esws/KollmerBWAK16}, where the \kgs allow to find semantic similarities between different domains;
\textbf{interpretable/explainable-recommendation}~\cite{DBLP:series/ssw/AnelliBNS20,9143460,DBLP:journals/kbs/YangD20,DBLP:conf/aaai/WangWX00C19,DBLP:conf/semweb/AnelliNSRT19}, with \kg being a backbone for understanding the recommendation model and providing human-like explanations;
\textbf{user-modeling}~\cite{DBLP:journals/tois/WangZWZLXG19,DBLP:conf/sac/Ojino19,DBLP:conf/um/KallumadiH18,DBLP:conf/es/LuoXCB14}, since the resource descriptions can drive the construction of the user profile;
\textbf{graph-based recommendation}~\cite{DBLP:journals/eswa/SangXQW21,DBLP:journals/access/WangSWXX20,DBLP:journals/kbs/ShiWXX20,DBLP:conf/www/WangZXLG19,DBLP:conf/cikm/WangZWZLXG18,DBLP:journals/tist/NoiaOTS16}, where the topology-based techniques have met the semantics of the edges/relations, and the ontological classification of nodes (classes);
\textbf{the cold-start problem}~\cite{DBLP:journals/fgcs/SahuD20,DBLP:journals/mis/YadavDB20,DBLP:journals/eswa/NatarajanVNG20,DBLP:journals/umuai/Fernandez-Tobias19}, since the \kgs can overcome the lack of collaborative information;
\textbf{the content-based recommendation}~\cite{DBLP:conf/recsys/AnelliNLS17,DBLP:conf/i-semantics/NoiaMORZ12} that solely relies on \kg and still produces high-quality recommendations.
\framework could be considered a Knowledge-aware hybrid collaborative/content-based recommendation model. While recent models of the same kind made use of Knowledge graph embeddings or factorization models, \framework considerably differs from them since it introduces the sparse factorization approach and reweights the user-feature interactions by exploiting the information gain signal. To the best of our knowledge, it is one of the first approaches to adopt this hybrid solution to obtain a personalized view of the embedding matrix.

\subsection{Entropy-driven Recommender Systems}
Entropy-based measures have been widely employed in recommendation systems.
A popular strategy to include entropy into the recommendation algorithm is to exploit it in connection with a similarity measure.
In this respect, \citet{DBLP:journals/ijis/WangZL15} proposed a new information entropy-driven user similarity-based model.
They suggest measuring the relative difference between ratings and develop a Manhattan distance-based model.
\citet{DBLP:journals/eswa/YalcinIB21} proposes two novel aggregation techniques by hybridizing additive utilitarian and approval voting methods to feature popular items on which group members provided a consensus. They use entropy to analyze rating distributions and detect items on which group members have reached no or little consensus. 
Entropy has also been used to model the purchase probability for a given set of recommendations for a specific user~\cite{DBLP:conf/www/IwataSY07}.
The idea is to exploit the maximum entropy principle by analyzing features in the recommendations and user interests.
Another example of the exploitation of entropy is~\citet{DBLP:journals/jaihc/Lee20}, where
they improve the previous similarity measures by employing the information entropy of user ratings to reflect the user’s global rating behavior on items.
\citet{DBLP:journals/umuai/KarimiNS15} proposed an innovative approach for active learning in recommender systems, aiming to take advantage of additional information.
They suggest employing entropy to drive the active learning process and increase the system performance for new users.
Entropy has also been applied to evaluate the quality and helpfulness of different product reviews~\cite{DBLP:journals/kais/ZhangT11}.
They propose an information gain-based model to predict the helpfulness of online product reviews to suggest the most suitable products and vendors to consumers.
Another interesting study~\cite{DBLP:conf/semweb/BouzaRBG08} integrates entropy more deeply into the recommendation process.
They calculate for every feature its information gain by considering item instances that provide the feature and item instances that do not.
Despite superficial similarities with~\citet{DBLP:conf/semweb/BouzaRBG08}, the two works are fundamentally different. In fact,~\citeauthor{DBLP:conf/semweb/BouzaRBG08} uses the class of the features with the highest information gain as a decision tree node, while \framework exploits the information gain to weigh the single user-feature interactions. Moreover, user and feature embeddings are combined using a dot-product similarity, showing some similarities with the former works. However, that is where the similarities end, since all the mentioned works propose completely different models from \framework (e.g., distance-based or active-learning models, feature selection techniques).

\section{Approach}
In the following, we introduce \framework. It exploits the knowledge encoded in a knowledge graph as side information to compute feature-aware user profiles, which are eventually used to provide personalized recommendation lists.

\subsection{Knowledge Graph and multi-hop predicates}
The Semantic Web was initially conceived to connect documents in the Web and improve data retrieving and access. 
Over the years, a full stack of semantic technologies emerged, leading to the Linking Open Data initiative~\cite{DBLP:series/synthesis/2011Heath}. The initiative indicates the remarkable effort of a community of researchers and practitioners to build publicly available knowledge bases of semantically linked machine-understandable data~\cite{bernerslee2001semantic}.
 Thanks to the Linked Data initiative, today, we can benefit from $1,483$ different \kgs connected in the so-called Linked Open Data Cloud\footnote{\url{https://lod-cloud.net/datasets}}.
These \kgs share the same ontology and the same schema across multiple domains, giving access to a wide-spread knowledge at the same development cost required for a single domain.
The most appreciated \kgs of this special class undoubtedly are DBpedia~\cite{DBLP:journals/semweb/LehmannIJJKMHMK15,DBLP:conf/semweb/AuerBKLCI07}, Wikidata~\cite{DBLP:journals/cacm/VrandecicK14,DBLP:conf/www/Vrandecic12}, Yago~\cite{DBLP:conf/www/SuchanekKW07} (the 4th release~\cite{DBLP:conf/esws/TanonWS20} also supports RDF*~\cite{DBLP:conf/amw/Hartig17}), FreeBase~\cite{DBLP:conf/sigmod/BollackerEPST08}, Satori\footnote{\url{https://searchengineland.com/library/bing/bing-satori}}\footnote{\url{https://blogs.bing.com/search/2013/03/21/understand-your-world-with-bing}}~\cite{DBLP:conf/cikm/LiuBLZSWX19,DBLP:journals/oir/UyarA15}, NELL~\cite{DBLP:conf/wsdm/CarlsonBWHM10}, Google's Knowledge Graph\footnote{\url{https://blog.google/products/search/introducing-knowledge-graph-things-not/}}, Facebook's Entities Graph\footnote{\url{https://www.facebook.com/notes/facebookengineering/under-the-hood-the-entitiesgraph/10151490531588920/}}, Knowledge Vault~\cite{DBLP:conf/kdd/0001GHHLMSSZ14}, Bio2RDF~\cite{DBLP:journals/jbi/BelleauNTRM08}.
This availability of \kgs is a clear advantage for KaRS.

A knowledge graph $\mathcal{KG}$ can be represented as a set of triples where entities are linked to each other by binary relations.
Each connection in $\mathcal{KG}$ is then a triple $\sigma \xrightarrow{\rho} \omega$, where $\sigma$ is a subject entity, $\rho$ is a relation (predicate), and $\omega$ is an object entity. Therefore, in $\mathcal{KG}$, the edge $\rho$ connects the entity $\sigma$ and the entity $\omega$ with a directed relation. Hereinafter, we generalize the previous notion to multi-hop predicates (i.e., considering chains of predicates that connect two entities at a higher depth). Let $n$-hop predicate be defined as $\rho = \langle \rho_1, ..., \rho_n \rangle$ if $\sigma \xrightarrow{\rho_1} \omega_1 \xrightarrow{\rho_2} ... \xrightarrow{\rho_n} \omega_n \in \mathcal{KG}$. For convenience, $h(\rho) = n$ for $\rho: \sigma \xrightarrow{\rho} \omega_n \in \mathcal{KG}$ denotes the depth of the predicate chain. When no confusion arises, from now on we will use $\sigma \xrightarrow{\rho} \omega$ to denote a generic chain with $h(\rho) \in \{1, ..., n\}$.

\subsection{Item and User Features in \framework}

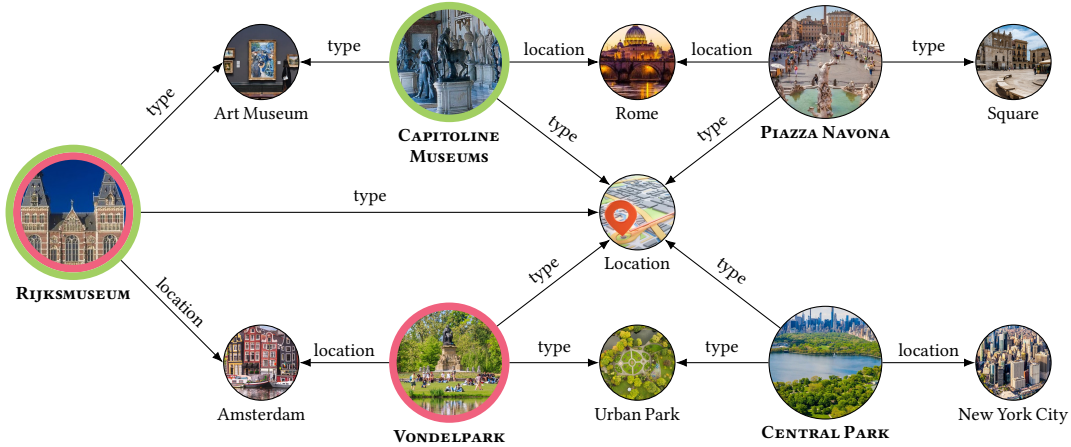
\begin{figure*}
    \centering
    \footnotesize
    \begin{tikzpicture}
    \node[draw=WildStrawberry!80,line width=1mm,circle,
    path picture={
               \node at (path picture bounding box.center){ \includegraphics[width=2cm]{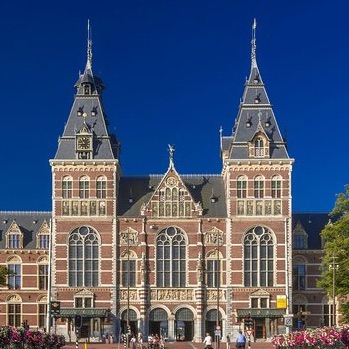}
               };
           },minimum width=1.5cm] (rm) at (0,0)  {};
    \node[draw=LimeGreen!80,line width=1mm,circle,minimum width=1.7cm,label=below:\textbf{\textsc{Rijksmuseum}}] (rm2) at (0,0)  {};

    \node[draw=LimeGreen!80,line width=1mm,circle,label={[text width=1.5cm, align=center]below:\textbf{\textsc{Capitoline\\Museums}}},
    path picture={
               \node at (path picture bounding box.center){ \includegraphics[width=2cm]{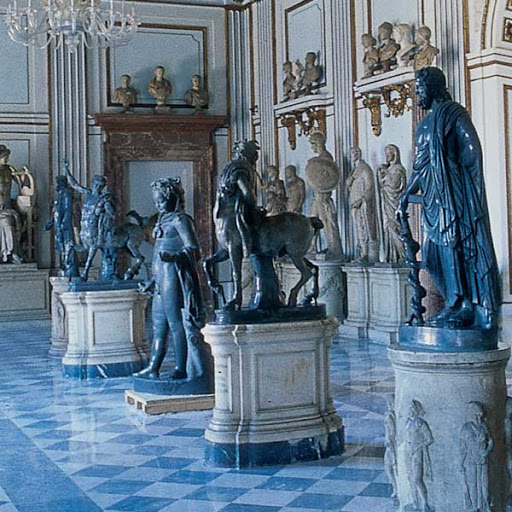}
               };
           },minimum width=1.5cm] (mc) at (5,2)  {};
    \node[draw,circle,label=below:\textbf{\textsc{Piazza Navona}},
    path picture={
               \node at (path picture bounding box.center){ \includegraphics[width=2cm]{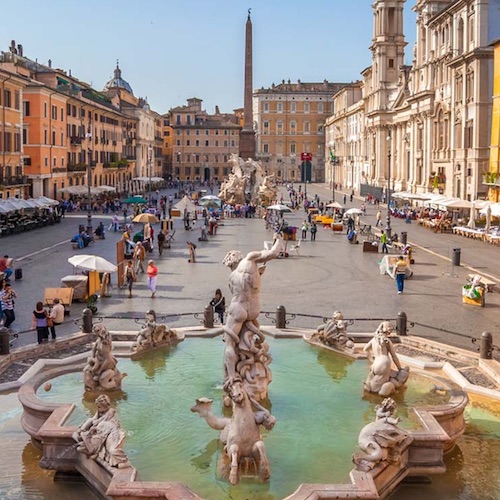}
               };
           },minimum width=1.5cm] (pn) at (10,2)  {};
    \node[draw=WildStrawberry!80,line width=1mm,circle,label=below:\textbf{\textsc{Vondelpark}},
    path picture={
               \node at (path picture bounding box.center){ \includegraphics[width=2cm]{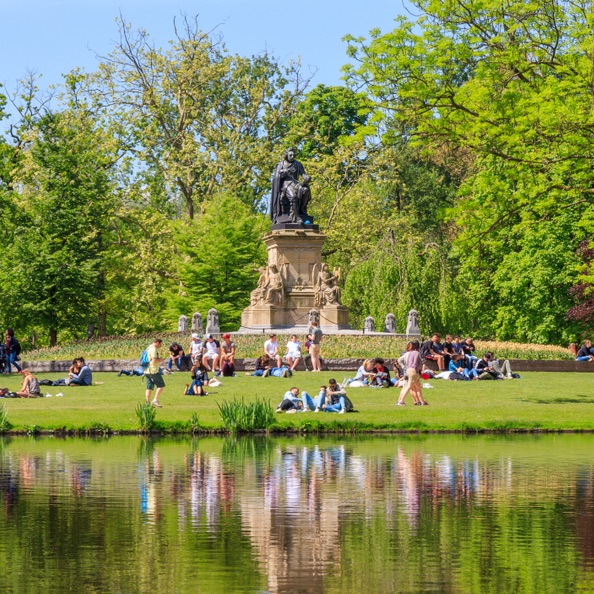}
               };
           },minimum width=1.5cm] (vp) at (5,-2)  {};
           
    \node[draw,circle,label=below:\textbf{\textsc{Central Park}},
    path picture={
               \node at (path picture bounding box.center){ \includegraphics[width=2cm]{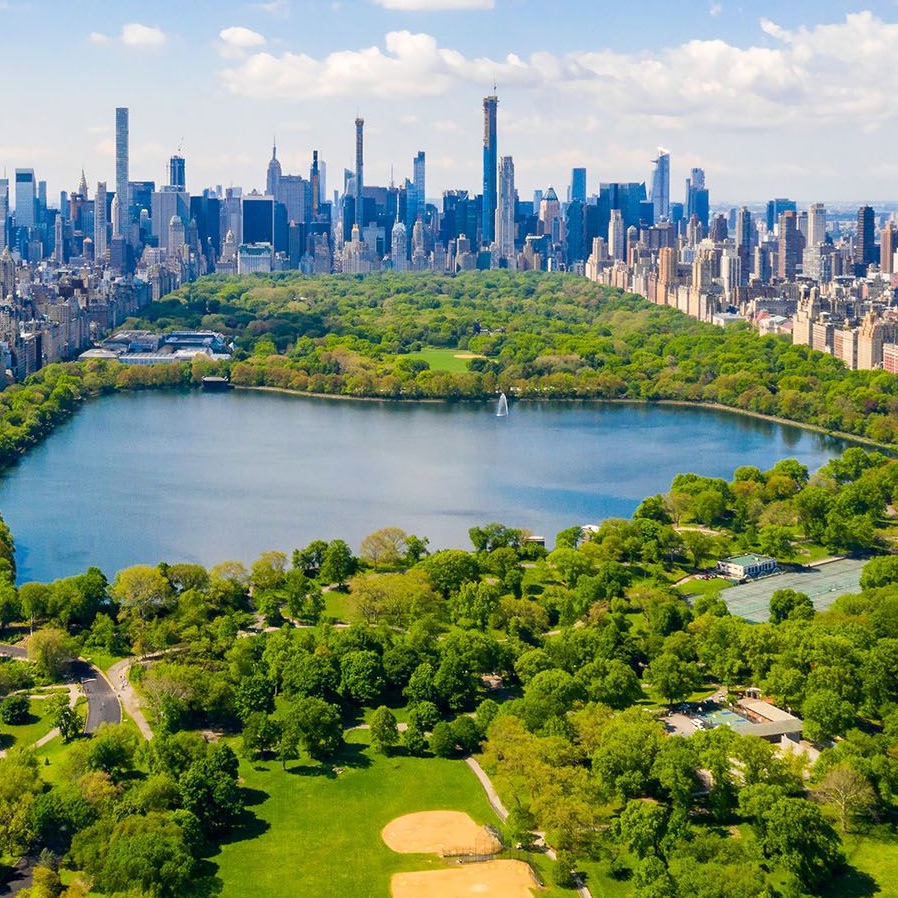}
               };
           },minimum width=1.5cm] (cp) at (10,-2)  {};

    \node[draw,circle,label=below:Amsterdam,
    path picture={
               \node at (path picture bounding box.center){ \includegraphics[width=1.5cm]{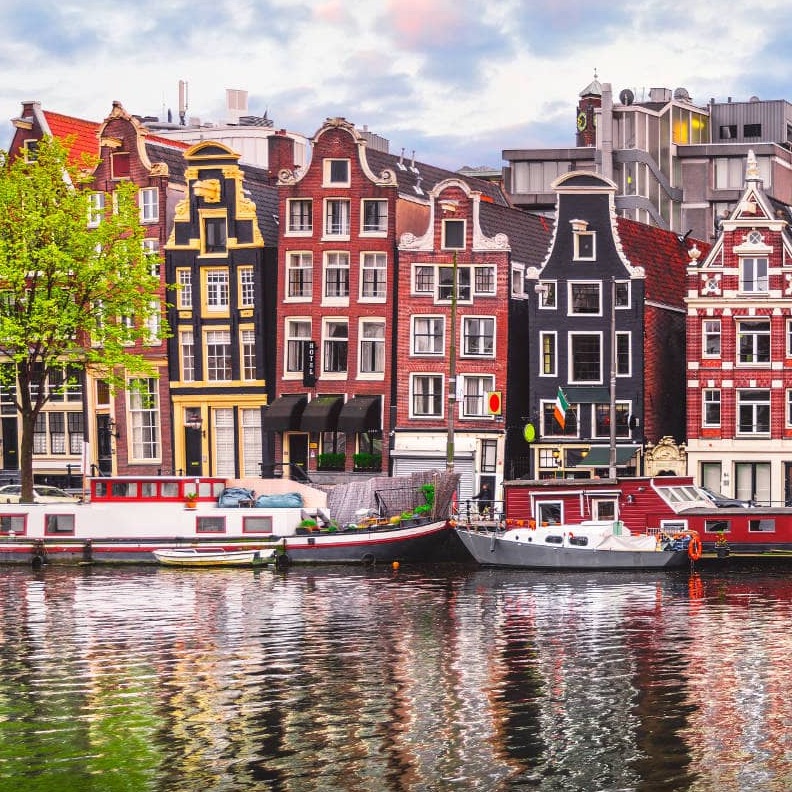}
               };
           },minimum width=1cm] (amst) at (2.5,-2)  {};

    \node[draw,circle,label=below:Art Museum,
    path picture={
               \node at (path picture bounding box.center){ \includegraphics[width=1.5cm]{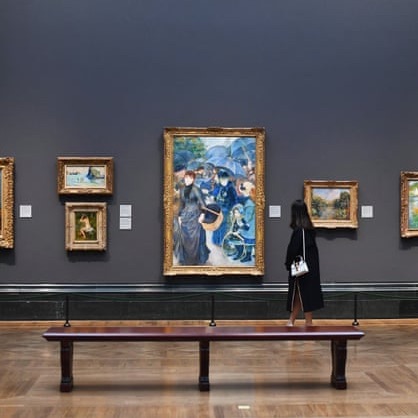}
               };
           },minimum width=1cm] (art) at (2.5,2)  {};

    \node[draw,circle,label=below:Location,
    path picture={
               \node at (path picture bounding box.center){ \includegraphics[width=1cm]{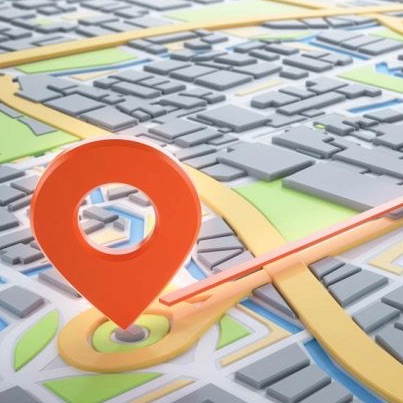}
               };
           },minimum width=1cm] (loc) at (7.5,0)  {};

    \node[draw,circle,label=below:Urban Park,
    path picture={
               \node at (path picture bounding box.center){ \includegraphics[width=1.5cm]{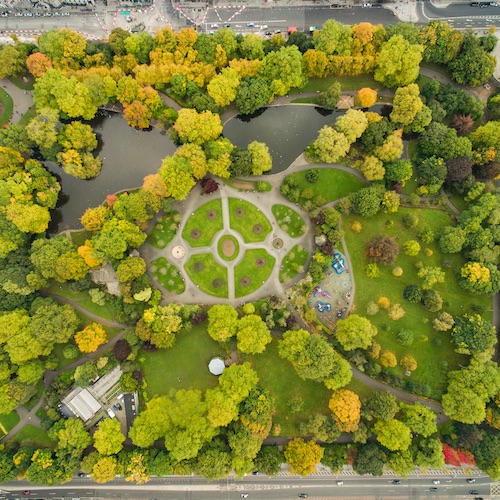}
               };
           },minimum width=1cm] (up) at (7.5,-2)  {};

    \node[draw,circle,label=below:New York City,
    path picture={
               \node at (path picture bounding box.center){ \includegraphics[width=1.5cm]{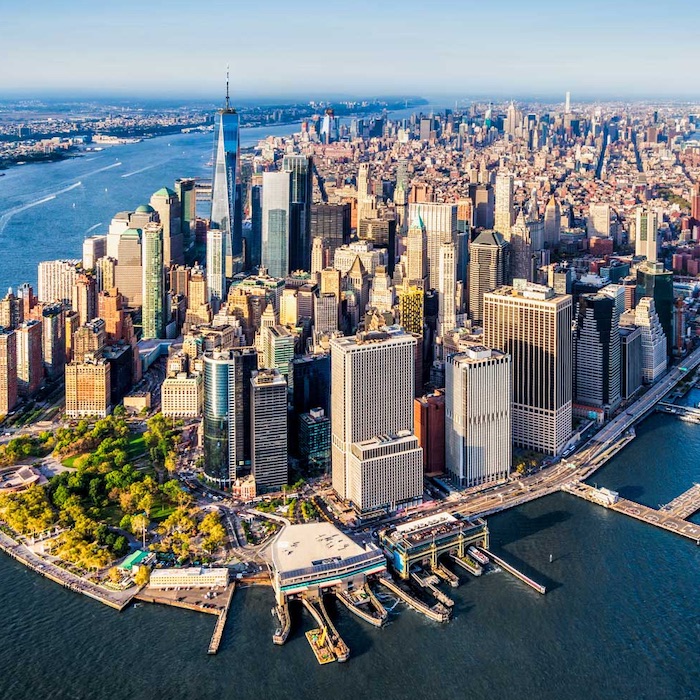}
               };
           },minimum width=1cm] (ny) at (12.5,-2)  {};

    \node[draw,circle,label=below:Square,
    path picture={
               \node at (path picture bounding box.center){ \includegraphics[width=1.5cm]{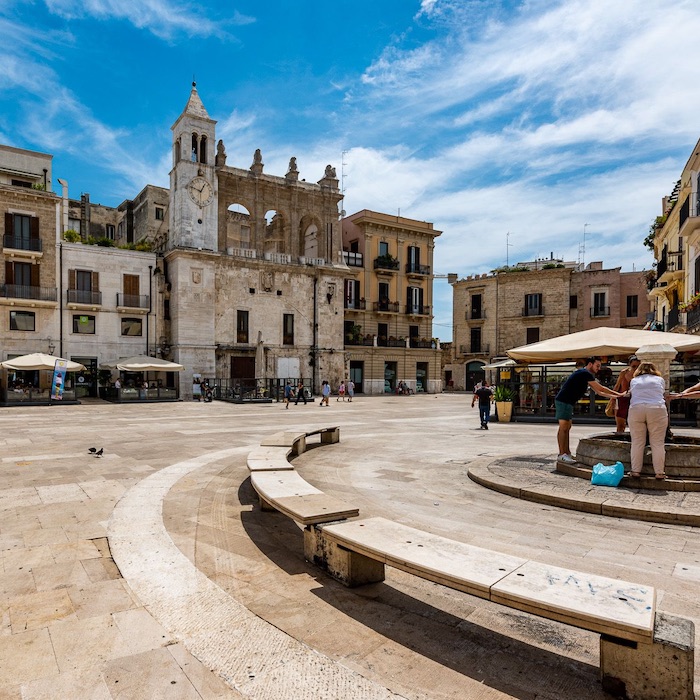}
               };
           },minimum width=1cm] (sq) at (12.5,2)  {};

    \node[draw,circle,label=below:Rome,
    path picture={
               \node at (path picture bounding box.center){ \includegraphics[width=1.5cm]{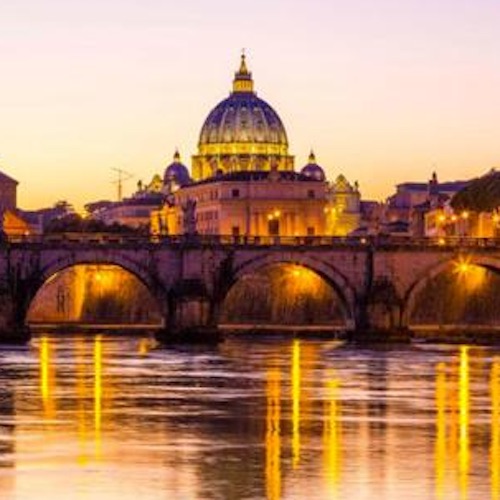}
               };
           },minimum width=1cm] (rome) at (7.5,2)  {};



    \draw[-{Latex[length=1.5mm]}] (rm2) -- node[above,sloped]{location} (amst.west);
    \draw[-{Latex[length=1.5mm]}] (rm2) -- node[above,sloped]{type} (art.west);
    \draw[-{Latex[length=1.5mm]}] (rm2) -- node[above,sloped]{type} (loc.west);
    
    \draw[-{Latex[length=1.5mm]}] (mc) -- node[above,sloped]{type} (loc.north west);
    \draw[-{Latex[length=1.5mm]}] (mc) -- node[above,sloped]{type} (art.east);
    \draw[-{Latex[length=1.5mm]}] (mc) -- node[above,sloped]{location} (rome.west);

    \draw[-{Latex[length=1.5mm]}] (pn) -- node[above,sloped]{location} (rome.east);
    \draw[-{Latex[length=1.5mm]}] (pn) -- node[above,sloped]{type} (loc.north east);
    \draw[-{Latex[length=1.5mm]}] (pn) -- node[above,sloped]{type} (sq.west);

    \draw[-{Latex[length=1.5mm]}] (vp) -- node[above,sloped]{location} (amst.east);
    \draw[-{Latex[length=1.5mm]}] (vp) -- node[above,sloped]{type} (loc.south west);
    \draw[-{Latex[length=1.5mm]}] (vp) -- node[above,sloped]{type} (up.west);

    \draw[-{Latex[length=1.5mm]}] (cp) -- node[above,sloped]{location} (ny.west);
    \draw[-{Latex[length=1.5mm]}] (cp) -- node[above,sloped]{type} (loc.south east);
    \draw[-{Latex[length=1.5mm]}] (cp) -- node[above,sloped]{type} (up.east);


    \end{tikzpicture}
    \caption{An excerpt of a knowledge graph, showing items of a catalog (Rijksmuseum, Vondelpark, Capitoline Museums, Piazza Navona, and Central Park) connected to other entities by predicates. Two users, Pink and Green, have expressed positive feedback for the items highlighted with the colors of their names.}
    \label{fig:kg_example}
    \vspace{-1em}
\end{figure*}

Given a collection of items $\mathcal{I}$ and a knowledge graph $\mathcal{KG}$ we assume each element in $i \in \mathcal{I}$ has a mapping to a corresponding entity in $\mathcal{KG}$. Under this assumption, an item $i$ can be explored, at depth $n$, to identify the set $\mathcal{F}_i^{(n)}$ of the semantic features describing it:
\begin{equation}
    \label{eq:feature-expl}
    \mathcal{F}_i^{(n)} = \{ \langle \rho, \omega\rangle  \;|\; i \xrightarrow{\rho} \omega \in \mathcal{KG}\;, h(\rho) \in \{1, ..., n\}\}.
\end{equation}
Once the features are extracted, \framework handles them equally, regardless of their original depth.

\begin{myexp*}
As an example, consider the \kg subgraph in Figure~\ref{fig:kg_example}, where a $1$-depth exploration has been performed for each of five items taken from a point-of-interest catalog. 
The formal \textit{Vondelpark} item description is:
\begin{align*}
\begin{autobreak}
\MoveEqLeft
\mathcal{F}^{(1)}_{\textit{Vondelpark}} = \{ \feature{type, Location},
\feature{location, Amsterdam}, \feature{type, Urban Park}\}.
\end{autobreak}
\tag*{$\square$}
\end{align*}
\end{myexp*}

We describe each user $u \in \mathcal{U}$ with a set $\mathcal{F}_u$ of features representing the items $\mathcal{I}_u \subseteq \mathcal{I}$ enjoyed by $u$.
We define $\mathcal{F}_u$ as the set of features describing the items that user $u$ has interacted with: 
\begin{equation}
   \mathcal{F}_u^{(n)} = \bigcup_{i\in \mathcal{I}_u} \mathcal{F}_i^{(n)}.
\end{equation}

\begin{rexp}
In Figure~\ref{fig:kg_example}, we represented the items enjoyed by two users by marking them with two different colors (green and pink).
To build the set $\mathcal{F}_\textit{Pink}^{(1)}$, the features of all the items appreciated by Pink have to be considered:
\begin{align*}
\begin{autobreak}
\MoveEqLeft
\mathcal{F}_\textit{Pink}^{(1)} = \{ \feature{type, Location},
\feature{location, Amsterdam},
\feature{type, Urban Park}, \feature{type, Art Museum} \}.
\end{autobreak}
\tag*{$\square$}
\end{align*}
\end{rexp}

Finally, the overall set $\mathcal{F}^{(n)}$ of the features in the system is defined as:
\begin{equation}
    \mathcal{F}^{(n)} = \bigcup_{i \in \mathcal{I}} \mathcal{F}_i^{(n)},
\end{equation}
with $\mathcal{F}_i^{(n)} \subseteq \mathcal{F}^{(n)}$ and $\mathcal{F}_u^{(n)} \subseteq \mathcal{F}^{(n)}$.
Depending on the value of $n$ and on the size of $\mathcal{I}$, the size of $\mathcal{F}^{(n)}$ could rapidly increase. 
Thus, filtering the item features might be a reasonable choice to control the computational and memory load and to improve the system performance.
Even though the literature about feature selection is vast, it is worth noticing that with \framework also graph pruning and semantic feature selection techniques~\cite{DBLP:conf/esws/NoiaMMPR18} could apply.  
In the following, for convenience, the $(n)$ superscript is omitted whenever it is not relevant in the context.

\subsection{Entropy of User Features}
\label{sec:entropy}
The main assumption behind \framework is that users make decisions (i.e., items to enjoy) based on a subset of item characteristics.
The assumption implies that not all the item features are equally important.
With \framework we move a step ahead in this direction by exploring how likely a user considers a feature in her item choice process.
Taking a cue from information theory, \framework exploits the notion of information gain to measure the relevance of a feature for a user in the process of deciding to consume or not the item.
For completeness, information gain is not the only metric used to select the best variable to partition data samples with respect to an outcome variable~\cite{DBLP:journals/tsmc/RokachM05}.
Nevertheless, information gain is widely adopted in a myriad of methods since it works also with non-binary values of each attribute. Moreover, for what regards the decision trees, the various informative metrics are quite consistent with each other and choosing one or another has a limited impact on the performance~\cite{DBLP:journals/amai/RaileanuS04}. For specific reasons, some metrics could be preferred (e.g., Gini impurity is well suited for its low computational cost~\cite{DBLP:journals/amai/RaileanuS04}).
However, a discussion about the advantages of the different metrics remains beyond the scope of this work.
In information theory, entropy is used to measure the uncertainty of a random variable. 
The entropy $H(V)$ of a random variable $V$ with $k$ possible values in $\{v_1, ..., v_k\}$ is defined as:  
\begin{equation}
\label{eq:entropy}
    H(V) = - \sum_{i=1}^k P(V = v_i) \log_2P(V = v_i).
\end{equation}
It is straightforward to check that a coin that always comes up heads has \textit{zero} entropy, while a fair coin equally likely to come up heads or tails when flipped has entropy $1$. Notably, if $V$ is a binary random variable that is true with probability $q$, we have $H(V) = B(q) = -(q\log_2q + (1-q)\log_2(1-q))$.
Therefore, given a dataset $\mathcal{D}$ of training samples in the form $(\mathbf{x}, y)$, with $\mathbf{x} \in \mathbb{R}^F$ and $y \in \{0, 1\}$, the entropy of the dataset is equal to $H(\mathcal{D})=B(P(y=1))$.

In this context, the information gain measures the expected reduction in information entropy from a prior state to a new state that acquires some information. With reference to the dataset $\mathcal{D}$, the new information comes from the observation of one of the attributes $x_d$ in $\mathbf{x}$. The $k$ distinct values $\{x_{d,1}, ... x_{d,k}\}$ that $x_d$ can assume partition the dataset $\mathcal{D}$ into $k$ mutually exclusive subsets, thus inducing a categorical probability distribution on the values of $x_d$. This gives the possibility to measure the expected entropy of $\mathcal{D}$ conditioned on $x_d$:
\begin{equation}
\label{eq:cond-entr}
    H(\mathcal{D}|x_d) = \sum_{i=1}^k P(x_d = x_{d,i}) H(\mathcal{D} | x_d = x_{d,i}).
\end{equation}
Then, we define the information gain $IG(\mathcal{D},x_d)$ obtained from the observation of the attribute $x_d$ as:
\begin{equation}
\label{eq:infogain}
	IG(\mathcal{D},x_d) = H(\mathcal{D}) - H(\mathcal{D}|x_d).
\end{equation}

The information gain defined in Eq. \eqref{eq:infogain} returns a measure of the importance of a single attribute in distinguishing positive from negative examples in a dataset. In \framework, we use the notion of information gain to measure how relevant a feature is to a user for deciding to consume or discard an item. In detail, to associate each feature of the system with an information gain, \framework uses the workflow described in the following.

For each user $u$, a dataset $\mathcal{D}_u$ is built with all the positive items (i.e., the items the user has enjoyed) from $\mathcal{I}_u$ and the same amount of negative items randomly picked up from $\bigcup_{v \in \mathcal{U}, v \neq u}\mathcal{I}_v \setminus \mathcal{I}_u$ (i.e., items not enjoyed by the user $u$ but enjoyed by other users). Therefore, following Eq.~\eqref{eq:entropy}, $H(\mathcal{D}_u) = 1$. Each sample is provided with a set of binary variables corresponding to the features in $\mathcal{F}_u$. Each variable indicates, for each item $i$ in $\mathcal{D}_u$, the presence ($f=1$) or the absence ($f=0$) of the corresponding feature in the set $\mathcal{F}_i$.

The information gain for each feature $f \in \mathcal{F}_u$ can be computed using the dataset $\mathcal{D}_u$.
Let $p_{uf}$ be the number of positive samples in $\mathcal{D}_u$ for which $f=1$, $n_{uf}$ the number of negative samples for which the same feature is present, and $t_{uf}$ the short form of $p_{uf} + n_{uf}$. Analogously, we define $p_{u \neg f} = |\mathcal{I}_u| - p_{uf}$ as the number of positive samples with $f = 0$,  $n_{u \neg f} = |\mathcal{I}_u| - n_{uf}$ as the number of negative samples with $f = 0$, and $t_{u\neg f}$ as the short form of $p_{u\neg f} + n_{u\neg f}$. Following Eqs.~\eqref{eq:cond-entr} and~\eqref{eq:infogain}:
\begin{equation}
\label{eq:info-gain}
    IG(\mathcal{D}_u, f) = 1 - H(\mathcal{D}_u | f=1) -  H(\mathcal{D}_u | f=0),
\end{equation}
\begin{equation}
H(\mathcal{D}_u | f=1) = \frac{t_{uf}}{|\mathcal{D}_u|} \bigg(- \frac{p_{uf}}{t_{uf}} \log_2\frac{p_{uf}}{t_{uf}} - \frac{n_{uf}}{t_{uf}} \log_2\frac{n_{uf}}{t_{uf}} \bigg),
\end{equation}
\begin{equation}
H(\mathcal{D}_u | f=0) = \frac{t_{u\neg f}}{|\mathcal{D}_u|} \bigg( - \frac{p_{u \neg f}}{t_{u\neg f}}\log_2\frac{p_{u \neg f}}{t_{u\neg f}} - \frac{n_{u \neg f}}{t_{u\neg f}} \log_2\frac{n_{u \neg f}}{t_{u\neg f}} \bigg),
\end{equation}
where $IG(\mathcal{D}_u, f)$ can be also merely regarded as a function in the values of $p_{uf}$ and $n_{uf}$.

In \framework, we associate a weight $k_{uf} = IG(\mathcal{D}_u,f)$ to each pair of user $u$ and feature $f$. It represents the influence of a feature ---in the view of the user--- in the prediction of user-item interactions. To compute the $k_{uf}$ values, the system designer can consider the whole set of features in $\mathcal{F}_u$, or filter them out according to a cutoff value of IG.

\begin{rexp}
To clarify the use of information gain in \framework and show its effects, we consider the example in Figure~\ref{fig:kg_example}. 
We see that Pink has visited the Rijksmuseum and the Vondelpark, both in Amsterdam. Thus \framework supposes she has a preference for the Dutch city. 
On the other hand, all the items in the catalog share the feature $\feature{type, Location}$, thus \framework assumes this latter not to be influential in the user decision-making process.
To build $\mathcal{D}_{\textit{Pink}}$, \framework combines the set of items experienced by Pink with a set of the same size containing items that Pink did not enjoy, e.g., $\mathcal{D}_{\textit{Pink}} = \{ \textit{Rijksmusem}, \textit{Vondelpark}, \textit{Piazza Navona}, \textit{Central Park}\}$. Let us observe the feature $\feature{location, Amsterdam}$. According to the previous definitions, it has to be influent for Pink. Given $\mathcal{D}_{\textit{Pink}}$, \framework computes:
\begin{align*}
\begin{autobreak}
\MoveEqLeft
p_{\textit{Pink}, \langle \textit{location}, \textit{Amsterdam} \rangle} = 2, \;\;\; n_{\textit{Pink}, \langle \textit{location}, \textit{Amsterdam} \rangle} = 0,
p_{\textit{Pink}, \neg \langle \textit{location}, \textit{Amsterdam} \rangle} = 0, \;\;\; n_{\textit{Pink}, \neg \langle \textit{location}, \textit{Amsterdam} \rangle} = 2.
\end{autobreak}
\end{align*}
Consequently, according to Eq.~\eqref{eq:info-gain}, $k_{\textit{Pink}, \langle \textit{location}, \textit{Amsterdam} \rangle} = 1$, meaning that Pink strongly takes into account if a place to visit is located in Amsterdam or not. 
Therefore, \framework considers this feature to have a high impact on generating the recommendations for Pink. 
Moreover, it could be observed that $k_{\textit{Pink}, \langle \textit{type}, \textit{Art Museum} \rangle} \approx 0.31$.
Since it is not completely clear how influential this feature is in Pink’s decisions, it will have a smaller influence on the predictions for Pink.  
Finally, $k_{\textit{Pink}, \langle \textit{type}, \textit{Location} \rangle}$ and $k_{\textit{Pink}, \langle \textit{type}, \textit{Urban Park} \rangle}$ have zero information gain and no influence on the predictions. 
In detail, the former is common to all the items and does not bring additional information. The latter is shared by the same number (i.e., one) of positive and negative samples in $\mathcal{D}_{\textit{Pink}}$.
So, it makes this feature useless in distinguishing positive from negative places for Pink.
\qed
\end{rexp}

\subsection{Model Architecture}
\framework does not contain explicit representations for users and items. 
Instead, it represents the features in $\mathcal{F}$ as embeddings in a latent space $\mathbb{R}^E$.
Moreover, \framework promotes the idea of having user fine-tuned versions of the same model~\cite{DBLP:conf/ecir/AnelliDNFN21,DBLP:conf/sac/AnelliDNFN21,DBLP:conf/aiia/AnelliDNF19}.
Therefore, in addition to a global latent representation of each feature in $\mathcal{F}$, it builds, for each user $u$, a personal view of each feature in $\mathcal{F}_u \subseteq \mathcal{F}$.
Each user combines the global embeddings with her personal feature representation to estimate the overall user-item interaction.

The recommendation model is structured into two distinct parts, both containing a dense representation of dimensionality $E$ for each feature $f \in \mathcal{F}$. 
On the one hand, \framework keeps a set $\mathcal{G}$ of global trainable embeddings and biases shared among all the users (see Section \ref{sec:l2r}):
\begin{equation}
	\mathcal{G} = \{(\mathbf{g}_f \in \mathbb{R}^E, b_f \in \mathbb{R}) \; | \; f \in \mathcal{F} \}.
\end{equation}
On the other hand, each user in \framework also has her personal representation of the features she interacted with, i.e., the features in $\mathcal{F}_u$. These embeddings are collected within the set $\mathcal{P}^u$, defined as:
\begin{equation}
	\mathcal{P}^u = \{\mathbf{p}^u_f \in \mathbb{R}^E \; | \; f \in \mathcal{F}_u \}.
\end{equation}

\framework estimates the possible affinity of a feature $f$ to the user $u$ with the inner product between the personal representation $\mathbf{p}^u_f$ and the global representation $\mathbf{g}_f$, plus the global bias term $b_f$. 
To estimate the overall affinity of user $u$ to item $i$, \framework combines the features shared by $u$ and $i$ and then weighs them according to their pre-computed entropy-based values.
In detail, being $\mathcal{F}_{ui} = \mathcal{F}_u \cap \mathcal{F}_i$ the set of common features between user $u$ and item $i$, \framework predicts their interaction $\hat{x}_{ui}$ as it follows:
\begin{equation}
    \label{eq:ui_interaction}
	\hat{x}_{ui} = \sum_{f \in \mathcal{F}_{ui}} k_{uf} (\mathbf{p}_f^u \mathbf{g}_f + b_f).
\end{equation}
Eq.~\eqref{eq:ui_interaction} encodes the strategy \framework exploits to handle thousands of model features. In fact, it takes advantage of the user profile to involve only a small subset of them in the estimate of the user-item affinity.

\begin{rexp}
From the previous analysis, it is clear how Pink's choices (see Figure~\ref{fig:kg_example}) are influenced by the features $\feature{location, Amsterdam}$ and $\feature{type, Art Museum}$. Consider that the interaction of Pink with \textit{Capitoline Museum} is to be estimated by \framework. The set of the common features is $\mathcal{F}_{\textit{Pink}, \textit{Capitoline Museum}} = \{ \langle \textit{type}, \textit{Art Museum} \rangle, \linebreak[1] \langle \textit{type}, \textit{Location} \rangle \}$. Accordingly to Eq.~\eqref{eq:ui_interaction}, the interaction is modelled as the summation of contributions of the common features:
\begin{align*}
\begin{autobreak}
\MoveEqLeft
\hat{x}_{\textit{Pink}, \textit{Capitoline Museum}} =
k_{Pink, \langle \textit{type}, \textit{Art Museum} \rangle} (\mathbf{p}^{Pink}_{\langle \textit{type}, \textit{Art Museum} \rangle} \mathbf{g}_{\langle \textit{type}, \textit{Art Museum} \rangle} + 
b_{\langle \textit{type}, \textit{Art Museum} \rangle}) +
k_{Pink, \langle \textit{type}, \textit{Location} \rangle}
(\mathbf{p}^{Pink}_{\langle \textit{type}, \textit{Location} \rangle} \mathbf{g}_{\langle \textit{type}, \textit{Location} \rangle} + 
b_{\langle \textit{type}, \textit{Location} \rangle}).
\end{autobreak}
\end{align*}
As expected, there is no contribution of the feature $\feature{type, Location}$ because of the value of its pre-computed entropy-based weight. Thus, the only contribution to the estimation is given by the embeddings of $\feature{type, Art Museum}$.
\qed
\end{rexp}

\subsection{Learning to Rank}
\label{sec:l2r}

To learn the model parameters, \framework adopts the well-known Bayesian Personalized Ranking (BPR) optimization criterion, which is a maximum posterior estimator for personalized ranking and the most common pair-wise Learning to Rank strategy.
BPR assumes that a user $u$ prefers a consumed item $i^+$ over a non-consumed item $i^-$, and optimizes the model by maximizing, for each pair of $i^+$ and $i^-$, a function of the difference $x_{ui^+} - x_{ui^-}$.
To update the model, \framework uses stochastic gradient descent, considered that the partial derivatives of the generic $\hat{x}_{ui}$ with respect to the model parameters are:
\begin{gather}
    \frac{\partial}{\partial \theta}\hat{x}_{ui} =
\begin{cases} 
k_{uf}\mathbf{g}_{f} & \mbox{if } \theta=\mathbf{p}^u_f, \\
k_{uf}\mathbf{p}^u_f & \mbox{if } \theta=\mathbf{g}_{f}, \\
k_{uf} & \mbox{if } \theta=b_f,\\
0 & \mbox{else.}
\end{cases}
\end{gather}

\section{Experiments}
\label{sec:exp-setup}
This section describes the design of the experimental setting, the evaluation protocol, and the baselines. First of all we introduce the Research Questions that drove the experimental evaluation of \framework:
\begin{description}
    \item[RQ1] Is \framework able to provide both accurate and diverse recommendation to users?
    \item[RQ2] Is \framework robust to popularity bias and disinclined to introduce algorithmic bias?
    \item[RQ3] What happens if \framework is deprived of high-order features?
    Which are the effects on accuracy and diversity?
    \item[RQ4] Do \framework recommendations preserve the semantics included in the original features?
\end{description}

\subsection{Datasets}
\label{sec:datasets}

We have evaluated the performance of \framework on three datasets from different domains, namely \textit{Yahoo! Movies}, \textit{MovieLens}, and \textit{Facebook Books}. Each item in these datasets is provided with a \textit{DBpedia} URI, which links to the semantic description of the item as an entity of the \textit{DBpedia} knowledge graph.
\textit{Yahoo! Movies} contains $69,846$ movie ratings generated on \textit{Yahoo! Movies} up to November $2003$ on a $[1, 5]$ scale. The ratings have been collected from $4,000$ users with respect to $2,626$ items.
It provides mappings to \textit{MovieLens} and \textit{EachMovie} datasets. We binarize the explicit data by keeping ratings of $3$ or higher and interpret them as positive implicit feedback.
The dataset \textit{MovieLens} is a collection of users' ratings in the movie domain: it contains $1,000,209$ ratings on a $[1, 5]$ scale from $6,040$ users with respect to $3,706$ items. Similar to \textit{Yahoo! Movies}, we binarize the explicit data by keeping ratings of $3$ or higher.
Finally, \textit{Facebook Books} is a more sparse dataset with $18,978$ positive implicit feedback from $1,398$ users about $2,933$ books. 
To ensure a fair comparison with the baselines, we applied an iterative $10$-core preprocessing on \textit{Yahoo! Movies} and \textit{MovieLens}, and a $5$-core preprocessing on \textit{Facebook Books}.

\subsection{Feature Extraction}
\label{sec:feature-extr}

For a fair comparison, we have
used the features resulting from the following workflow for \framework and for the baselines that make use of content information, i.e., VSM and kaHFM.

\noindent\textbf{Exploration of Knowledge Graph.}
The items of the datasets have been described with a set of semantic features retrieved through a knowledge graph exploration at depth 2 in the form of $\langle \rho, \omega \rangle$ pairs (see Eq.~\ref{eq:feature-expl}).
The semantic information has been retrieved from the \textit{DBpedia} knowledge graph\footnote{\url{https://www.dbpedia.org}}, thanks to the item-to-\textit{DBpedia}-URI mapping provided with the datasets. Some features (based on their 1-hop predicate) have not been considered, since they provide auxiliar information not useful for characterizing the content of the item \cite{DBLP:conf/esws/NoiaMMPR18}. In detail, we filtered out the predicates \textit{dbo:wikiPageWikiLink}, 
\textit{owl:sameAs},
\textit{rdf:type},
\textit{gold:hypernym},
\textit{rdfs:seeAlso}, \textit{dbp:wordnet\_type},
\textit{dbo:wikiPageExternalLink}, 
\textit{dbo:thumbnail}, \textit{prov:wasDerivedFrom}, and  
\textit{dbp:wikiPageUsesTemplate}.

\noindent\textbf{Feature Filtering based on Frequency.} 
Irrelevant features have been removed due to the poor information they bring and to reduce the computational costs. Thus, we have removed the features that are common to less than 10 items. The resulting features constitute the common set of features for all the competing models that make use of content features, although some of these models ---including \framework--- may operate some further filtering operations.

\noindent\textbf{Feature Filtering based on Entropy in \framework. }
As mentioned in Section~\ref{sec:entropy}, features in the user personal representation have been filtered as well based on their information gain. For instance, in a 2-hop exploration of the knowledge graph, users with a large number of transactions could reach an impressive number of nodes of the knowledge graph. Thus, we filtered out features with little information gain keeping, as a maximum limit for each user, the 100 most informative features from the 1-hop exploration and the 100 most informative features from the 2-hop exploration. 

\subsection{Baselines}
To assess the effectiveness of \framework, we compare it with various baselines. In particular, we are interested in comparing \framework with other latent factor models and other state-of-the-art baselines which helps to position the model with respect to some of the best recommendation approaches in the literature.

\noindent\textbf{Non-competing Algorithms. } Random and Most Popular are two non-personalized recommenders used for reference. Among content-based algorithms, we have chosen the Vector Space Model (VSM)~\cite{DBLP:conf/i-semantics/NoiaMORZ12} where user and item profiles have been generated with TF-IDF and cosine similarities between them have been computed. 
As collaborative-filtering baselines, we have considered Item-kNN ~\cite{DBLP:journals/tkdd/Koren10} (an item-based implementation of the k-nearest neighbors algorithm) and MultiVAE~\cite{DBLP:conf/www/LiangKHJ18}, a non-linear probabilistic model taking advantage of Bayesian inference to estimate the parameters.

\noindent\textbf{Competing Algorithms. } We compare \framework against  factorization-based algorithms, both non-neural and neural. Among them, we consider i) BPR-MF~\cite{DBLP:conf/uai/RendleFGS09}, a latent factor model based on the same pair-wise optimization criterion used in \framework, ii) the neural Matrix Factorization in the version of~\citet{DBLP:conf/recsys/RendleKZA20}, iii) NeuMF~\cite{DBLP:conf/sigir/0001C17}, and iv) kaHFM~\cite{DBLP:conf/semweb/AnelliNSRT19}, another factorization-based model making use of knowledge graphs for model building and initialization.

\subsection{Reproducibility, Evaluation Protocol and Metrics}
We have chosen the \textit{all unrated items} protocol to compare the  different algorithms. In \textit{all unrated items}, for each user we consider as recommendable items all the items not yet rated by that user. We have split the datasets using hold-out 80-20 splitting strategy, retaining for each user the 80\% of her ratings in the training set and the remaining 20\% in the test set~\cite{DBLP:reference/sp/GunawardanaS15}.
All the models have been tested in 10 different configurations of hyperparameters, according to the Bayesian hyperparameter optimization algorithm. For the sake of reproducibility, we provide our code and a working configuration file for the framework Elliot~\cite{DBLP:conf/sigir/AnelliBFMMPDN21}, with complete and ready-to-use information about the experiments we have run.
We have measured the recommendation accuracy by exploiting nDCG ~\cite{DBLP:conf/kdd/KricheneR20}. It has been also used for validation and choosing the best hyperparameter configurations.
We have also evaluated the diversity of recommendation, adopting Item Coverage ~\cite{DBLP:journals/tkde/AdomaviciusK12} and Gini Index ~\cite{DBLP:reference/sp/CastellsHV15} (higher is better). The former provides the overall number of diverse recommended items, and it highlights the degree of personalization. The latter measures how unequally a system provides users with different items, with higher values corresponding to more tailored lists.
Finally, three bias metrics have been used to evaluate how \framework and the baselines behave on the underrepresentation of items from the long-tail. To this aim we have used ACLT (higher is better), which measures the fraction of the long-tail items the recommender has covered~\cite{DBLP:conf/flairs/AbdollahpouriBM19}. Moreover, we have also evaluated PopREO and PopRSP (smaller is better, in $[0,1]$), which are specific applications of RSP and REO~\cite{DBLP:conf/sigir/ZhuWC20}. Notably, PopREO estimates the equal opportunity of items, encouraging the true positive rate of popular and unpopular items to be same. PopRSP is a measure of statistical parity, assessing whether the ranking probability distributions for popular and unpopular items are the same in recommendation.

\section{Discussion}
\begin{table*}[tbp]
\caption{Comparison of \framework with competing baselines (names in boldface) and other reference baselines on Yahoo! Movies, Facebook Books, and MovieLens 1M. The best result among the competing baselines is in boldface, the second-best result is underlined.}
\small
\setlength{\tabcolsep}{0.75em}
\renewcommand{\arraystretch}{0.9}
\vspace{-0.5em}
\begin{tabular}{lrrrrrrrrrrrr}

\hlineB{2.5}
\myrowcolour
 \multicolumn{13}{c}{\textbf{a) Yahoo! Movies}} \\

\toprule
 & \multicolumn{ 2}{c}{\textbf{nDCG}} & \multicolumn{ 2}{c}{\textbf{Item Coverage}} & \multicolumn{ 2}{c}{\textbf{Gini Index}} & \multicolumn{ 2}{c}{\textbf{ACLT}} & \multicolumn{ 2}{c}{\textbf{PopREO}} & \multicolumn{ 2}{c}{\textbf{PopRSP}} \\
 \cmidrule(r){2-3} \cmidrule(lr){4-5} \cmidrule(lr){6-7}
 \cmidrule(lr){8-9} \cmidrule(lr){10-11} \cmidrule(l){12-13}
 & @10 & @1 & @10 & @1 & @10 & @1 & @10 & @1 & @10 & @1 & @10 & @1\\
  \cmidrule(r){2-2} \cmidrule(lr){3-3}
  \cmidrule(lr){4-4} \cmidrule(lr){5-5} \cmidrule(lr){6-6} \cmidrule(lr){7-7} \cmidrule(lr){8-8} \cmidrule(lr){9-9} \cmidrule(lr){10-10} \cmidrule(lr){11-11} \cmidrule(lr){12-12} \cmidrule(l){13-13}
Random & 0.00960 & 0.00842 & 1050 & 811 & 0.84956 & 0.56591 & 5.52026 & 0.54727 & 0.09847 & 0.53357 & 0.00980 & 0.00341 \\ 
Most Popular & 0.15850 & 0.13666 & 49 & 11 & 0.01263 & 0.00103 & 0.00000 & 0.00000 & 1.00000 & 1.00000 & 1.00000 & 1.00000 \\ 
VSM & 0.04777 & 0.04534 & 370 & 93 & 0.05245 & 0.01388 & 3.11768 & 0.22701 & 0.49588 & 0.54432 & 0.45751 & 0.61122 \\ 
Item-kNN & 0.30739 & 0.34715 & 745 & 297 & 0.15826 & 0.09853 & 0.98842 & 0.08617 & 0.70585 & 0.70651 & 0.83466 & 0.85618 \\ 
MultiVAE & 0.23696 & 0.24547 & 399 & 152 & 0.09136 & 0.04187 & 0.23473 & 0.01222 & 0.85433 & 0.82931 & 0.96127 & 0.97988 \\ 
\textbf{BPR-MF} & 0.18571 & 0.17098 & 151 & 35 & 0.02191 & 0.00412 & 0.00064 & 0.00000 & 0.99543 & 1.00000 & 0.99989 & 1.00000 \\ 
\textbf{MF} & \underline{0.28971} & 0.29987 & 455 & 177 & 0.09024 & 0.04640 & 0.08232 & 0.00257 & 0.87345 & 0.93531 & 0.98645 & 0.99577 \\ 
\textbf{NeuMF} & 0.09184 & 0.08549 & 50 & 13 & 0.01134 & 0.00094 & 0.00064 & 0.00000 & 1.00000 & 1.00000 & 0.99989 & 1.00000 \\ 
\textbf{kaHFM} & \textbf{0.30055} & \textbf{0.32383} & \underline{757} & \underline{290} & \underline{0.16591} & \underline{0.09875} & \underline{0.46238} & \underline{0.03344} & \underline{0.76103} & \underline{0.74940} & \underline{0.92339} & \underline{0.94472} \\ 
\textbf{\framework} & 0.24640 & \underline{0.31218} & \textbf{851} & \textbf{370} & \textbf{0.28015} & \textbf{0.13612} & \textbf{2.14469} & \textbf{0.11447} & \textbf{0.44768} & \textbf{0.62916} & \textbf{0.63355} & \textbf{0.80797} \\ 
\bottomrule
\end{tabular}

\vspace{0.4em}

\begin{tabular}{lrrrrrrrrrrrr}

\hlineB{2.5}
\myrowcolour
 \multicolumn{13}{c}{\textbf{b) Facebook Books}} \\

\toprule
 & \multicolumn{ 2}{c}{\textbf{nDCG}} & \multicolumn{ 2}{c}{\textbf{Item Coverage}} & \multicolumn{ 2}{c}{\textbf{Gini Index}} & \multicolumn{ 2}{c}{\textbf{ACLT}} & \multicolumn{ 2}{c}{\textbf{PopREO}} & \multicolumn{ 2}{c}{\textbf{PopRSP}} \\
 \cmidrule(r){2-3} \cmidrule(lr){4-5} \cmidrule(lr){6-7}
 \cmidrule(lr){8-9} \cmidrule(lr){10-11} \cmidrule(l){12-13}
 & @10 & @1 & @10 & @1 & @10 & @1 & @10 & @1 & @10 & @1 & @10 & @1\\
  \cmidrule(r){2-2} \cmidrule(lr){3-3}
  \cmidrule(lr){4-4} \cmidrule(lr){5-5} \cmidrule(lr){6-6} \cmidrule(lr){7-7} \cmidrule(lr){8-8} \cmidrule(lr){9-9} \cmidrule(lr){10-10} \cmidrule(lr){11-11} \cmidrule(lr){12-12} \cmidrule(l){13-13}
Random & 0.00690 & 0.00587 & 782 & 646 & 0.86167 & 0.58782 & 5.26045 & 0.53705 & 0.09794 & 0.11591 & 0.00749 & 0.00877 \\ 
Most Popular & 0.09393 & 0.08291 & 16 & 4 & 0.01265 & 0.00065 & 0.00000 & 0.00000 & 1.00000 & 1.00000 & 1.00000 & 1.00000 \\ 
VSM & 0.03617 & 0.02128 & 523 & 203 & 0.18874 & 0.08392 & 3.80558 & 0.30668 & 0.22616 & 0.76175 & 0.29958 & 0.44088 \\ 
Item-kNN & 0.12903 & 0.09244 & 769 & 338 & 0.37520 & 0.16060 & 2.22524 & 0.21056 & 0.48852 & 0.43001 & 0.59861 & 0.62075 \\ 
MultiVAE & 0.11914 & 0.08291 & 620 & 197 & 0.18279 & 0.06344 & 0.46368 & 0.02715 & 0.77695 & 0.87226 & 0.91818 & 0.95221 \\ 
\textbf{BPR-MF} & 0.09473 & 0.08291 & 17 & 4 & 0.01318 & 0.00066 & 0.00000 & 0.00000 & 1.00000 & 1.00000 & 1.00000 & 1.00000 \\ 
\textbf{MF} & \underline{0.09557} & \underline{0.08437} & 87 & 16 & 0.02376 & 0.00118 & 0.00000 & 0.00000 & 1.00000 & 1.00000 & 1.00000 & 1.00000 \\ 
\textbf{NeuMF} & 0.07142 & 0.07557 & 17 & 4 & 0.01245 & 0.00062 & 0.00000 & 0.00000 & 1.00000 & 1.00000 & 1.00000 & 1.00000 \\ 
\textbf{kaHFM} & \textbf{0.12667} & \textbf{0.08584} & \underline{540} & \underline{174} & \underline{0.13866} & \underline{0.06071} & \underline{0.32942} & \underline{0.02494} & \underline{0.87663} & \underline{0.93678} & \underline{0.94197} & \underline{0.95610} \\ 
\textbf{\framework} & 0.08526 & 0.06530 & \textbf{606} & \textbf{288} & \textbf{0.30703} & \textbf{0.15876} & \textbf{3.02641} & \textbf{0.24578} & \textbf{0.15210} & \textbf{0.13150} & \textbf{0.44852} & \textbf{0.55535} \\ 
\bottomrule
\end{tabular}

\vspace{0.4em}

\begin{tabular}{lrrrrrrrrrrrr}

\hlineB{2.5}
\myrowcolour
 \multicolumn{13}{c}{\textbf{c) MovieLens 1M}} \\

\toprule
 & \multicolumn{ 2}{c}{\textbf{nDCG}} & \multicolumn{ 2}{c}{\textbf{Item Coverage}} & \multicolumn{ 2}{c}{\textbf{Gini Index}} & \multicolumn{ 2}{c}{\textbf{ACLT}} & \multicolumn{ 2}{c}{\textbf{PopREO}} & \multicolumn{ 2}{c}{\textbf{PopRSP}} \\
 \cmidrule(r){2-3} \cmidrule(lr){4-5} \cmidrule(lr){6-7}
 \cmidrule(lr){8-9} \cmidrule(lr){10-11} \cmidrule(l){12-13}
 & @10 & @1 & @10 & @1 & @10 & @1 & @10 & @1 & @10 & @1 & @10 & @1\\
  \cmidrule(r){2-2} \cmidrule(lr){3-3}
  \cmidrule(lr){4-4} \cmidrule(lr){5-5} \cmidrule(lr){6-6} \cmidrule(lr){7-7} \cmidrule(lr){8-8} \cmidrule(lr){9-9} \cmidrule(lr){10-10} \cmidrule(lr){11-11} \cmidrule(lr){12-12} \cmidrule(l){13-13}
Random & 0.00961 & 0.00894 & 3203 & 2701 & 0.86573 & 0.60357 & 6.57351 & 0.65662 & 0.07153 & 0.04044 & 0.00560 & 0.00200 \\ 
Most Popular & 0.19836 & 0.25124 & 69 & 19 & 0.00540 & 0.00061 & 0.00000 & 0.00000 & 1.00000 & 1.00000 & 1.00000 & 1.00000 \\ 
VSM & 0.04761 & 0.03726 & 170 & 34 & 0.00618 & 0.00114 & 1.53328 & 0.36291 & 0.83004 & 0.34127 & 0.82635 & 0.53954 \\ 
Item-kNN & 0.36897 & 0.48029 & 980 & 395 & 0.05838 & 0.03180 & 0.06656 & 0.00414 & 0.96059 & 0.97073 & 0.99299 & 0.99565 \\ 
MultiVAE & 0.34244 & 0.40030 & 1794 & 856 & 0.13805 & 0.09207 & 0.44553 & 0.04139 & 0.76521 & 0.76053 & 0.95220 & 0.95566 \\ 
\textbf{BPR-MF} & \textbf{0.36755} & \textbf{0.46787} & \underline{1141} & \textbf{527} & \underline{0.07654} & \textbf{0.04491} & \underline{0.06325} & \underline{0.00182} & \underline{0.95507} & 0.99326 & \underline{0.99334} & \underline{0.99809} \\ 
\textbf{MF} & 0.18325 & 0.20785 & 100 & 35 & 0.00751 & 0.00292 & 0.00000 & 0.00000 & 1.00000 & 1.00000 & 1.00000 & 1.00000 \\ 
\textbf{NeuMF} & 0.13741 & 0.14955 & 70 & 18 & 0.00458 & 0.00040 & 0.00000 & 0.00000 & 1.00000 & 1.00000 & 1.00000 & 1.00000 \\ 
\textbf{kaHFM} & \underline{0.32218} & \underline{0.42034} & 955 & 337 & 0.04818 & 0.02193 & 0.04503 & 0.00166 & 0.97196 & \underline{0.98877} & 0.99526 & 0.99826 \\ 
\textbf{\framework} & 0.19820 & 0.27542 & 1403 & \underline{492} & \textbf{0.08436} & \underline{0.03271} & \textbf{0.86126} & \textbf{0.04801} & \textbf{0.82072} & \textbf{0.93842} & \textbf{0.90570} & \textbf{0.94841} \\ \bottomrule
\end{tabular}
\vspace{-1.5em}

\label{tbl:yahoo-general}
\end{table*}

In the following, we discuss the main insights coming from the performed experiments, with the aim of answering the Research Questions posed in Section~\ref{sec:exp-setup}.

\subsection{Accuracy and Diversity: an Analytical and Qualitative Study (RQ1)}

The first analysis aims to answer RQ1.
In fact, the purpose of this evaluation is to assess whether \framework is capable to provide accurate and diverse recommendation.
Tables~\ref{tbl:yahoo-general}a, \ref{tbl:yahoo-general}b and \ref{tbl:yahoo-general}c show an analysis of the accuracy and diversity performance comparing \framework with the other baselines in terms of nDCG and Gini Index. The best and the second-best values are highlighted with \textbf{boldface} and \underline{underline}, respectively. 
Although several baselines are considered (to position the methods), a thorough comparison is mainly made with respect to the latent factor models, highlighted in boldface in the first column of tables.
The results in Tables~\ref{tbl:yahoo-general}a, \ref{tbl:yahoo-general}b and \ref{tbl:yahoo-general}c have been statistically validated with Student Paired t-test and Wilcoxon test, with a $p$-value level of 0.05. The complete significance hypothesis test tables are available in the \framework repository.
The general behavior that can be observed at first glance from Table~\ref{tbl:yahoo-general}a is that \framework exhibits a satisfactory performance regarding the accuracy, being outperformed only by kaHFM and MF in the top-$10$ recommendation. Moreover, it behaves even better in the top-$1$ task, where its nDCG value becomes comparable to the accuracy result of kaHFM, which shows the best accuracy performance. \framework significantly outperforms BPR-MF, although both are learned with a pair-wise BPR optimization, suggesting the useful role of the extracted knowledge.
When looking at the diversity performance represented by the item coverage and Gini values, we note the high degree of personalization provided by \framework. We link this result to the personalized view of the knowledge granted by the framework. Moreover, in \framework the collaborative signal on explicit user interests ensures to recommend diverse items among the ones sharing characteristics of interest for the user.
The behavior pointed out so far is not entirely confirmed in Facebook Books (see Table~\ref{tbl:yahoo-general}b). Indeed, here the accuracy results remain below the performance of other factorization-based approaches. However, the diversity results show how BPR-MF, MF and NeuMF may have been completely flooded by popularity signal, which led them to perform poorly regarding the item coverage and Gini metrics. Instead, \framework approaches the superior performance of Item-kNN in terms of diversity, and even here it shows an improvement in top-$1$ recommendation, where the gap is reduced.
Furthermore, we analyze Table~\ref{tbl:yahoo-general}c, showing the performance on MovieLens 1M. BPR-MF performs superbly on this dataset, while the other factorization-based approaches ---but kaHFM--- remain significantly below its capabilities. This downward trend seems to be caused by the strong popularity signal, which is prevailing in MF and NeuMF. Instead, \framework does not suffer from this problem and it is the best model in terms of diversity, while providing still meaningful recommendations.

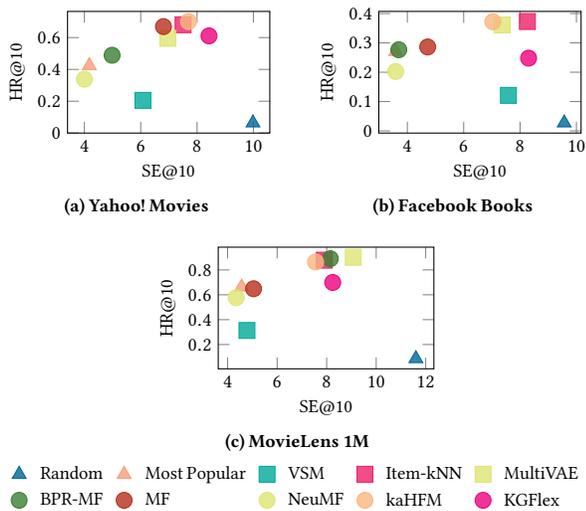
\begin{figure}%
    \footnotesize
    
    \centering
    \subfloat[Yahoo! Movies]{

\begin{tikzpicture}
    \begin{axis}[
            width=0.24*(\textwidth/\linewidth)*\linewidth,
            yticklabel style={
            /pgf/number format/fixed,
            /pgf/number format/precision=3
            },
            scaled y ticks=false,
            xlabel=SE@10,
            ylabel=HR@10,
            ylabel near ticks,
            xlabel near ticks,
            height=3.2cm
    ]

        \addplot[scatter, fill opacity=0.8,  mark size=3pt, only marks,
         scatter/classes={
            0={color=MidnightBlue, mark=triangle*},
            1={color=Melon, mark=triangle*},
            2={color=JungleGreen, mark=square*},
            3={color=WildStrawberry, mark=square*},
            4={color=GreenYellow, mark=square*},
            5={color=OliveGreen, mark=*},
            6={color=BrickRed, mark=*},
            7={color=GreenYellow, mark=*},
            8={color=Apricot, mark=*},
            9={color=RubineRed, mark=*}
        },
        scatter src=explicit symbolic] table [x=yahoo-se10, y=yahoo-hr10, col sep=comma, meta=class] {images/csv/kg_flex_acc_div.csv};

    \end{axis}
    \end{tikzpicture}

    }%
    \qquad
    \subfloat[Facebook Books]{

\begin{tikzpicture}
    \begin{axis}[
            width=0.24*(\textwidth/\linewidth)*\linewidth,
            yticklabel style={
            /pgf/number format/fixed,
            /pgf/number format/precision=3
            },
            scaled y ticks=false,
            xlabel=SE@10,
            ylabel=HR@10,
            ylabel near ticks,
            xlabel near ticks,
            height=3.2cm
    ]

    \addplot[scatter, fill opacity=0.8,  mark size=3pt, only marks,
     scatter/classes={
        0={color=MidnightBlue, mark=triangle*},
        1={color=Melon, mark=triangle*},
        2={color=JungleGreen, mark=square*},
        3={color=WildStrawberry, mark=square*},
        4={color=GreenYellow, mark=square*},
        5={color=OliveGreen, mark=*},
        6={color=BrickRed, mark=*},
        7={color=GreenYellow, mark=*},
        8={color=Apricot, mark=*},
        9={color=RubineRed, mark=*}
    },
    scatter src=explicit symbolic] table [x=fb-se10, y=fb-hr10, col sep=comma, meta=class] {images/csv/kg_flex_acc_div.csv};
    
    \end{axis}
    \end{tikzpicture}

    }%
    \qquad
    \subfloat[MovieLens 1M]{

\begin{tikzpicture}
    \begin{axis}[
            width=0.25*(\textwidth/\linewidth)*\linewidth,
            yticklabel style={
            /pgf/number format/fixed,
            /pgf/number format/precision=3
            },
            scaled y ticks=false,
            xlabel=SE@10,
            ylabel=HR@10,
            ylabel near ticks,
            xlabel near ticks,
            height=3.2cm,
    ]

    \addplot[scatter, fill opacity=0.8,  mark size=3pt, only marks,
     scatter/classes={
        0={color=MidnightBlue, mark=triangle*},
        1={color=Melon, mark=triangle*},
        2={color=JungleGreen, mark=square*},
        3={color=WildStrawberry, mark=square*},
        4={color=GreenYellow, mark=square*},
        5={color=OliveGreen, mark=*},
        6={color=BrickRed, mark=*},
        7={color=GreenYellow, mark=*},
        8={color=Apricot, mark=*},
        9={color=RubineRed, mark=*}
    },
    scatter src=explicit symbolic] table [x=ml-se10, y=ml-hr10, col sep=comma, meta=class] {images/csv/kg_flex_acc_div.csv};

    \end{axis}
    

    \end{tikzpicture}
    
    }%
    
    \begin{tikzpicture}
    \begin{customlegend}[legend columns=5,legend style={draw=none,column sep=1ex}, legend cell align={left},
            legend entries={Random, Most Popular, VSM, Item-kNN, MultiVAE, BPR-MF, MF, NeuMF, kaHFM, \framework}]
            \addlegendimage{fill opacity=0.8,  mark size=3pt, only marks,color=MidnightBlue, mark=triangle*}
            \addlegendimage{fill opacity=0.8,  mark size=3pt, only marks,color=Melon, mark=triangle*}
            \addlegendimage{fill opacity=0.8,  mark size=3pt, only marks,color=JungleGreen, mark=square*}
            \addlegendimage{fill opacity=0.8,  mark size=3pt, only marks,color=WildStrawberry, mark=square*}
            \addlegendimage{fill opacity=0.8,  mark size=3pt, only marks,color=GreenYellow, mark=square*}
            \addlegendimage{fill opacity=0.8,  mark size=3pt, only marks,color=OliveGreen, mark=*}
            \addlegendimage{fill opacity=0.8,  mark size=3pt, only marks,color=BrickRed, mark=*}
            \addlegendimage{fill opacity=0.8,  mark size=3pt, only marks,color=GreenYellow, mark=*}
            \addlegendimage{fill opacity=0.8,  mark size=3pt, only marks,color=Apricot, mark=*}
            \addlegendimage{fill opacity=0.8,  mark size=3pt, only marks,color=RubineRed, mark=*}
            \end{customlegend}
    \end{tikzpicture}
    
    \caption{Accuracy vs. distributional diversity. The plots show the value of HR@10 against SE@10: the closer to the top-right corner the better.}%
    \label{fig:acc-div}%
    \vspace{-2em}
\end{figure}

What we have analytically observed is confirmed in Figure~\ref{fig:acc-div}. These graphs show the joint behavior of \framework on accuracy and distributional diversity, by analyzing the value of Hit Ratio (HR) on the top-$10$ recommendation lists with respect to the Shannon Entropy (SE) statistics. Among factorization-based approaches (labelled in the plots), \framework approaches the right-top margin to a greater extent. The kaHFM model usually is the second-best model, but, on MovieLens 1M, BPR-MF shows its best performance. The other approaches seem to perform very poorly in at least one dimension or do not have a stable position when varying the dataset. This confirms the previous findings, and gives \framework the merit of providing highly personalized recommendations, thanks to the joint operation of the global and the personal views of the same features.

\subsection{Induction and Amplification of the Bias (RQ2)}

The behavior of \framework led us to analyze the quality of the recommendation in terms of
popularity bias,
a frequent problem causing popular items to be more and more recommended and less popular ones to remain underrepresented~\cite{DBLP:conf/recsys/AbdollahpouriMB19}. This algorithmic bias may cause a fairness issue from the item point of view, but also an inappropriate recommendation for users who do not prefer very popular items.
Tables~\ref{tbl:yahoo-general}a, \ref{tbl:yahoo-general}b and ~\ref{tbl:yahoo-general}c provide the values of three metrics to measure the bias. 
\framework always outperforms all the other factorization-based approaches and generally outperforms the other approaches. Regarding \framework, the Average Coverage of Long-Tail items, measured by ACLT (the higher the better), is comparable with the value obtained by VSM. This result is supported by the values of PopREO and PopRSP (the smaller the better), which encourage the ranking probability distributions and the true positive rates of popular and less popular items to be the same. Indeed, \framework and VSM grant the less biased recommendations. Interestingly, while both exploit the same optimization criterion, we notice how \framework consistently improves BPR-MF, which is known to be vulnerable to imbalanced data and to produce biased recommendations~\cite{DBLP:conf/sigir/ZhuWC20}.
To conclude, we can easily assert that the personalized representation of content information gives \framework the push to provide satisfactory and diverse recommendations without being negatively affected by popularity bias.

\subsection{The impact of Knowledge Graph exploration (RQ3)}

\begin{figure}%
    \footnotesize
    
    \centering
    \subfloat[Accuracy]{

\begin{tikzpicture}
    \begin{axis}[
        width  = 0.48*\linewidth,
        height = 3.8cm,
        major x tick style = transparent,
        ybar=2*\pgflinewidth,
        bar width=4pt,
        ymajorgrids = true,
        ylabel = {nDCG@10},
        symbolic x coords={KGFlex-0,KGFlex-1,KGFlex-2,KGFlex},
        xticklabels={\frameworkwithoutspace\textsuperscript{($\emptyset$)},\frameworkwithoutspace\textsuperscript{(1)},\frameworkwithoutspace\textsuperscript{(2)},\framework},
        xtick = data,
        scaled y ticks = false,
        enlarge x limits=0.2,
        ymin=0,
        yticklabel style={
        /pgf/number format/fixed,
        /pgf/number format/precision=3
        },
        scaled y ticks=false,
        ylabel near ticks,
        x tick label style={rotate=35,anchor=east}
    ]

        \addlegendimage{empty legend}

        \addplot[ style={MidnightBlue,fill=MidnightBlue,mark=none}, select coords between index={0}{3}] table [x=features, y=nDCG, col sep=comma] {images/csv/ablation_ml.csv};
        
        \addplot[style={LimeGreen,fill=LimeGreen,mark=none}, select coords between index={4}{7}] table [x=features, y=nDCG, col sep=comma] {images/csv/ablation_ml.csv};
        
        \addplot[ style={WildStrawberry,fill=WildStrawberry,mark=none}, select coords between index={8}{11}] table [x=features, y=nDCG, col sep=comma] {images/csv/ablation_ml.csv};


    \end{axis}
    \end{tikzpicture}

    }%
    \qquad
    \subfloat[Diversity]{
    
\begin{tikzpicture}
    \begin{axis}[
        width  = 0.48*\linewidth,
        height = 3.8cm,
        major x tick style = transparent,
        ybar=2*\pgflinewidth,
        bar width=4pt,
        ymajorgrids = true,
        ylabel = {Gini@10},
        symbolic x coords={KGFlex-0,KGFlex-1,KGFlex-2,KGFlex},
        xticklabels={\frameworkwithoutspace\textsuperscript{($\emptyset$)},\frameworkwithoutspace\textsuperscript{(1)},\frameworkwithoutspace\textsuperscript{(2)},\framework},
        xtick = data,
        scaled y ticks = false,
        enlarge x limits=0.2,
        ymin=0,
        yticklabel style={
        /pgf/number format/fixed,
        /pgf/number format/precision=3
        },
        scaled y ticks=false,
         ylabel near ticks,
                 x tick label style={rotate=35,anchor=east}
]

        \addplot[
        style={MidnightBlue,fill=MidnightBlue,mark=none}, select coords between index={0}{3}] table [x=features, y=Gini, col sep=comma] {images/csv/ablation_ml.csv};
        
        \addplot[
        style={LimeGreen,fill=LimeGreen,mark=none}, select coords between index={4}{7}] table [x=features, y=Gini, col sep=comma] {images/csv/ablation_ml.csv};
        
        \addplot[
        style={WildStrawberry,fill=WildStrawberry,mark=none}, select coords between index={8}{11}] table [x=features, y=Gini, col sep=comma] {images/csv/ablation_ml.csv};


    \end{axis}
    
    \vspace{-1.5em}

\end{tikzpicture}
    
    }%
    
    \begin{tikzpicture}
    \begin{customlegend}[legend columns=4,legend style={draw=none,column sep=1ex}, legend cell align={left},
            legend entries={\textbf{Embedding size}, \hspace{-.5em}1, \hspace{-.5em}10, \hspace{-.5em}100}]
            \addlegendimage{empty legend}            \addlegendimage{mark size=3pt, only marks,color=MidnightBlue, mark=*}
              \addlegendimage{mark size=3pt, only marks,color=LimeGreen, mark=*}
                \addlegendimage{mark size=3pt, only marks,color=WildStrawberry, mark=*}
            \end{customlegend}
    \end{tikzpicture}

    \caption{Ablation study on MovieLens 1M of \framework evaluated with respect of accuracy and diversity, which shows that features from the second hop significantly improves the model. \frameworkwithoutspace\textsuperscript{($\emptyset$)} does not use features from knowledge graph, \frameworkwithoutspace\textsuperscript{(1)} only uses 1-hop features, \frameworkwithoutspace\textsuperscript{(2)} only uses 2-hop features. The colors represent different embedding sizes used for training.}%
\label{fig:ablation}
\vspace{-1.5em}
\end{figure}
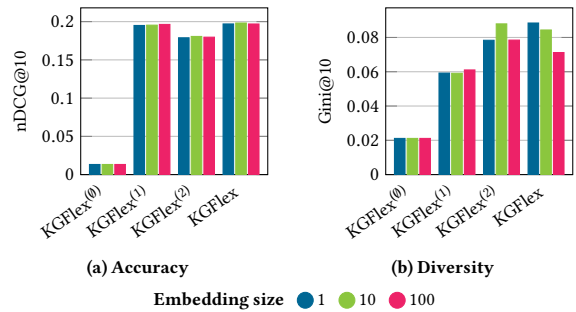

In the previous experimental setting (see Section~\ref{sec:feature-extr}), the personal user knowledge was represented by her $100$ most informative $1$-hop features and her $100$ most informative $2$-hop features. To give an intuition of how beneficial is the exploitation of these features in \framework, we performed an ablation study on MovieLens 1M in which we force \framework not to use features from the first hop exploration, or from the second hop exploration, or both.
Figure~\ref{fig:ablation} shows the accuracy and diversity performance of \framework, and its ablated versions \frameworkwithoutspace\textsuperscript{(1)}, using only features from the first hop, \frameworkwithoutspace\textsuperscript{(2)}, using only features from the second hop, and \frameworkwithoutspace\textsuperscript{($\emptyset$)}, which eliminates both. Moreover, we also plot how each version performs based on the embedding size, to understand whether it can affect the model variants.
As expected, \frameworkwithoutspace\textsuperscript{($\emptyset$)} performed in an unsatisfactory way on the recommendation task, since it cannot establish common content between users and items.
With 100 first-hop features per user (\frameworkwithoutspace\textsuperscript{(1)}), the system provides accurate recommendation, but its diversity performance remains low.
In this configuration, changing the embedding size is not beneficial neither for diversity nor for accuracy.
When exploiting only features from the second hop \frameworkwithoutspace\textsuperscript{(2)} the situation changes: these features enable \framework to catch more information about the content of the items and their relation, with a beneficial effect for the diversity of the recommendation.
The information carried by the second-hop features has more probability of embodying the actual reason why a user decides to enjoy an item (e.g., a user may watch a TV show not strictly for the director himself but rather for his nationality).
Finally, with the addition of the first-hop features, the complete version of \framework overcomes \frameworkwithoutspace\textsuperscript{(2)} in accuracy, regardless of the embedding size. This latter slightly penalizes the diversity, very likely due to increased awareness regarding item popularity.
This aspect requires further investigation and suggests room to increase the \framework performance further. The study definitely answers RQ3. As expected, the lack of a piece of knowledge negatively impacts the system. However, interestingly, the combination of first- and second-hop features positively impacts on accuracy and diversity performance of \framework.

\subsection{Preservation of the Original Semantics (RQ4)}

\begin{figure}%
    \centering
    \subfloat[Original dataset]{{\label{subig:a}\includegraphics[width=0.34\textwidth]{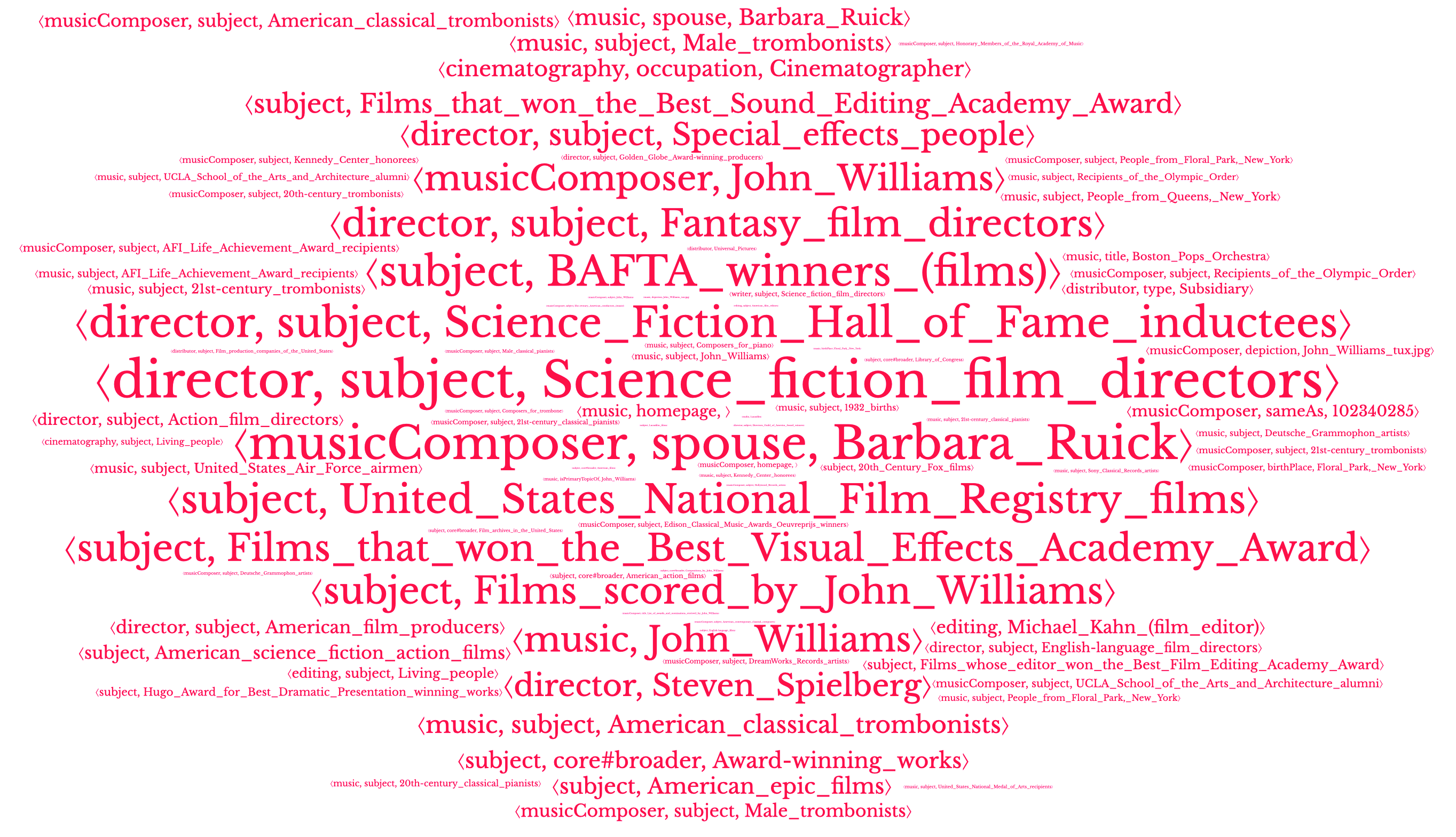} }}%
    \qquad
    \subfloat[Recommended lists]{{\label{subfig:b}\includegraphics[width=0.34\textwidth]{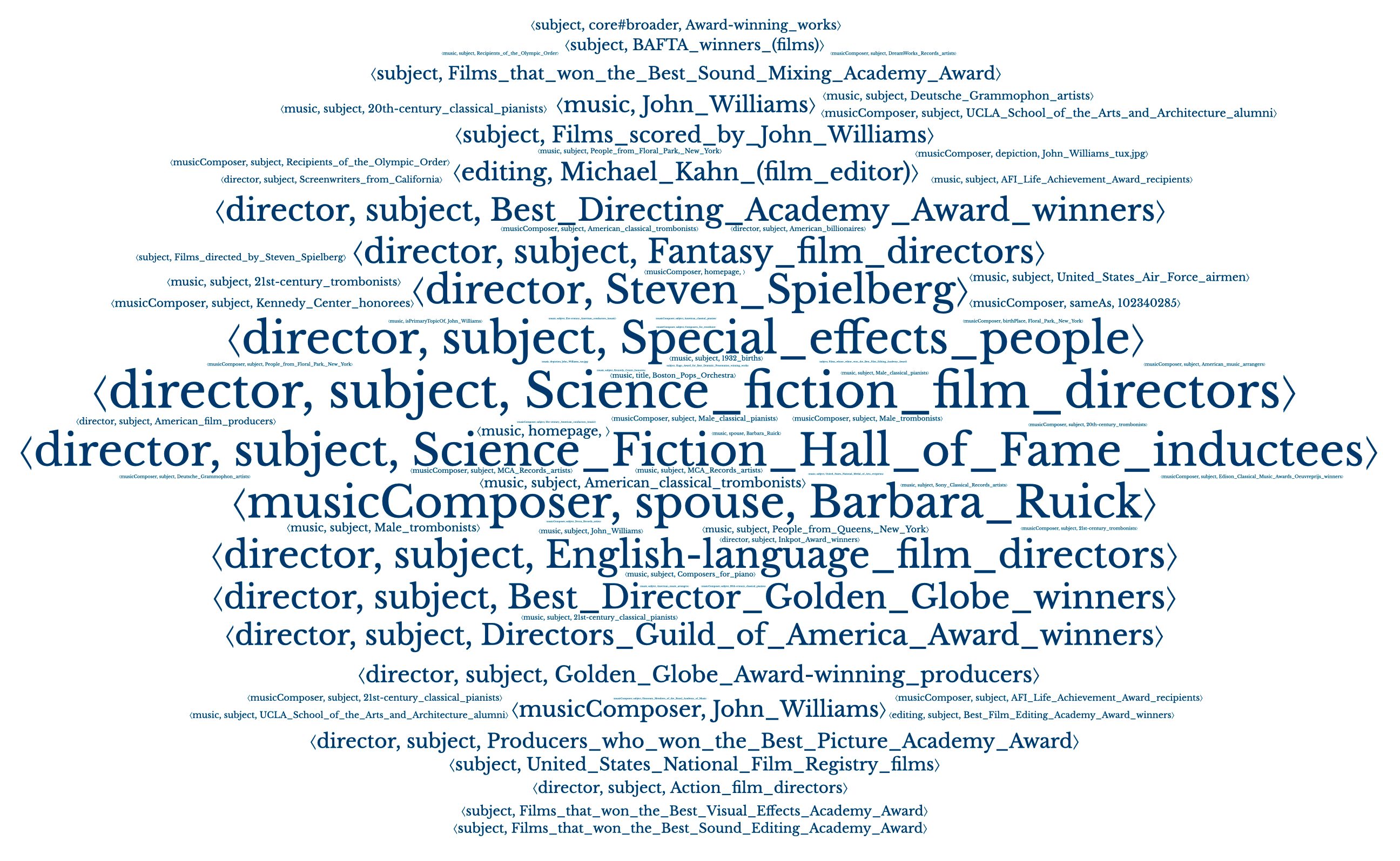}}}%
    \caption{Two word clouds showing the 100 most informative features for the users in Yahoo! Movies on the original dataset and on the recommendation lists. The word clouds suggest that \framework is able to preserve the original semantics included in the dataset.}%
    \label{fig:wordclouds}%
\end{figure}

\framework makes extensive use of side information and takes advantage of it in various training phases.
Nevertheless, there are no guarantees that \framework is really able to catch the semantic information encoded and use it to propose coherent recommendations.
To investigate this aspect, we have analyzed the features characterizing the recommended items and we have compared those features with the features derived from the user's historical interactions.
We have conducted a graphical experiment, showing the information gain of each feature in the system before and after the recommendation. Figure~\ref{fig:wordclouds} depicts the word clouds of the features that could be involved in user decision-making gathered from all the users in Yahoo! Movies. 
In detail, Figure~\ref{subig:a} represents the prominence of each feature, in terms of information gain, in the original dataset. 
Instead, in Figure~\ref{subfig:b}, the same analysis is performed on the recommendation lists provided by \framework.
For the sake of readability, both the word clouds visualize the 100 most informative features.
We observe that topics related to science fiction persist at the top, including the interest for the director Steven Spielberg and for John Williams (also by means of his spouse Barbara Ruick) and the interest in fantasy films award-winner movies. More precisely, $79\%$ of the top $100$ informative features from the dataset are associated with items in the recommendation list. 
Additionally, for each user, we have computed the percentage of her $k$ most informative features that have been retained in her recommendation list, with $k \in \{ 5, 10, 50, 100, \infty \}$. Table~\ref{tab:my_label} shows, for each column, the first, the second (median), and the third quartile of such percentages over all the population of users. It is remarkable how \framework provides a higher coverage of the original semantic when considering features more important to the users (i.e., lower $k$).
Finally, with the support of Figure~\ref{fig:wordclouds} and Table~\ref{tab:my_label}, we answer to RQ4. Overall, it could be easily observed that \framework preserves the main users' interests involved in decision-making. This suggests that \framework is able to preserve the original semantics, deeply integrating the content-based information into its recommendations.
Finally, this evidence suggests that we could be a step closer to providing users with items that they would have chosen autonomously.

\begin{table}[]
\small
    \caption{Percentage of preserved top-$k$ user features after the recommendation. The first ($Q_1$), the second ($Q_2$), and the third ($Q_3$) quartile (over all the population of users) of per-user percentages are shown.}
    \centering
    \renewcommand{\arraystretch}{0.9}
    \begin{tabular}{cccccc}
    \toprule
          & $k=\infty$ & $k=100$ & $k=50$ & $k=10$ & $k=5$\\
          \midrule
         \textbf{$Q_1$}& 14.5\% & 11\% & 12\% & 20\% & 20\%\\
         \textbf{$Q_2$} & 21.5\% & 21\% &26\% & 30\% & 40\%\\
         \textbf{$Q_3$} &31.5\% & 37\% & 48\% & 60\% & 60\%\\
         \bottomrule
    \end{tabular}
    \label{tab:my_label}
    \vspace{-1.5em}
\end{table}

\subsection{Limitations of \framework}

Section~\ref{sec:datasets} explicitly refers to datasets linking to a knowledge graph such as \textit{DBpedia}.
Nonetheless, the approach behind \framework works independently of the type of side information.
In fact, a dataset with each item linked to a generic set of features or attributes would suffice \framework. However, the quality of the side information may have a profound impact on the final performance of the method.
Moreover, if the side information is structured as a graph, also the depth of the exploration may impact the performance.
In this work, we exploited knowledge graphs as side information since they are recognized to provide high-quality information.
Nevertheless, their unavailability does not preclude the use of \framework.

\section{Conclusion}
This paper has introduced \framework for producing knowledge-aware recommendations from implicit feedback. \framework takes the best from content-based and factorization-based recommendation approaches for building a sparse model, where features extracted from a knowledge graph are embedded in a latent space. The interactions between users and items in \framework are combinations of users' feature representations and global feature representations, weighted according to the importance of each feature.
\framework showed its superior behavior in terms of item diversity on three datasets while being very accurate and resilient to algorithmic bias. We have also shown the role a knowledge graph may play in feeding \framework with side information and how the extracted features preserve their semantics in the recommendation lists.
This new method seems to be highly flexible and suited to practical applications. As future work, we plan to extend the approach with a finer feature selection, new types of feedback, alternatives to information gain, other types of side information, and other losses. Finally, we will further investigate feature embeddings to achieve an even more precise representation of features.

\begin{acks}
The authors acknowledge partial support of the projects PON ARS01\_00876 BIO-D
Casa delle Tecnologie Emergenti della Città di Matera, PON ARS01\_00821 FLET4.0,
PIA Servizi Locali 2.0 H2020 Passapartout - Grant n. 101016956, PIA ERP4.0, IPZS-PRJ4\_IA\_NORMATIVO.
\end{acks}

\bibliographystyle{ACM-Reference-Format}
\bibliography{bibliography}

\end{document}